\documentclass[amsmath,amssymb, aps,prb,superscriptaddress,twocolumn,floatfix]{revtex4-2}
\usepackage{graphicx}% Include figure files
\usepackage{dcolumn}% Align table columns on decimal point
\usepackage{bm}
\usepackage[usenames]{color}
\usepackage{xcolor}
\usepackage{soul}
\usepackage{comment}

\usepackage{hyperref} 
\usepackage{physics}
\usepackage{bookmark}
\usepackage{mathtools}
\usepackage{subfigure}
\usepackage{amsthm}
\newtheorem{theorem}{Theorem}
\newtheorem{lemma}{Lemma}

\newcommand{\xUZH}{Department of Physics, University of Z{\"u}rich, Winterthurerstrasse 190, 8057 Z{\"u}rich, Switzerland}
\newcommand{\xcornell}{Department of Physics, Cornell University, Ithaca, NY, USA}

\newcommand{\xPSI}{Condensed Matter Theory Group, PSI Center for Scientific Computing, Theory and Data, Paul Scherrer Institute, 5232 Villigen PSI, Switzerland}

\usepackage{wasysym}

\newcounter{para}
\newcommand{\para}{\par\refstepcounter{para}\textbf{{\color{cyan}[\thepara]}}\space}

\let\para\relax

\begin{document}
\title{Supernematic}

\author{Dan Mao}
\affiliation{\xcornell}
\affiliation{\xUZH}
\affiliation{\xPSI}
\author{Eun-Ah Kim}
\affiliation{\xcornell}
\affiliation{Department of Physics, Ewha Womans University, Seoul, South Korea}

\begin{abstract}
Quantum theory of geometrically frustrated systems is usually approached as a gauge theory where the local conservation law becomes the Gauss law. 
Here we show that it can do something fundamentally different: enforce a global conserved quantity via a non-perturbative tiling invariant, rigorously linking microscopic geometry to a new macroscopically phase-coherent state. 
In a frustrated bosonic model on the honeycomb lattice in the cluster-charging regime at fractional filling, this mechanism protects a conserved global quantum number, the sublattice polarization $\tilde{N} = N_A - N_B$. 
Quantum fluctuation drives the spontaneous symmetry breaking of this global $U(1)$ symmetry to result in
 a supernematic phase — an incompressible yet phase-coherent quantum state that breaks rotational symmetry without forming a superfluid or realizing topological order.
This establishes a route to a novel quantum many-body state driven by combinatorial constraints. The realization of the SN on programmable quantum platforms can test the ergodicity of the resonance operator in constrained quantum dynamics and explore the use of quantum hardware to tackle classically hard tiling problems.
\end{abstract}

\maketitle
\tableofcontents
\section{Introduction}
\para Quantum systems at weak and strong coupling are governed by opposing organizing principles: kinetic energy favors delocalized, wave-like motion, while interaction energy drives particles toward optimal spatial organization. Although the uncertainty principle complicates descriptions that bridge these regimes, each limit is often simple to characterize. In the weak-coupling limit, the system forms a delocalized momentum eigenstate. In the strong-coupling limit, interactions can stabilize a unique spatial arrangement through crystallization, as in the Wigner crystal\cite{Wigner} or the Mott insulator\cite{Mott_1949}. Yet when additional internal degrees of freedom or geometric constraints frustrate these interactions, the optimization problem admits a manifold of equally favorable configurations rather than a single solution. In such cases, even the classical limit becomes profoundly nontrivial, giving rise to emergent gauge structures that connect degenerate states, as exemplified by spin ice\cite{castelnovo2008magnetic} and dimer models\cite{moessner2001short}. Understanding how quantum fluctuations navigate this degenerate landscape—whether they select an ordered phase or preserve the imprint of frustration—remains one of the most conceptually challenging problems in modern quantum many-body physics.

\para The fate of geometric frustration under quantum fluctuations is often analyzed through gauge theories. The derivation of such theories are not analytically tractable in general, but perturbing away from exactly soluble models has been productive~\cite{RK,KITAEV20062}. When the gauge theory is deconfined, the resulting state could be topologically ordered \cite{Moessner2001resonating}. However, one may ask whether there is an alternative, rigorous path by which the intricate microscopic geometric constraints imprint their signature on a quantum ground state.
In this paper, we provide one such example. We will prove that geometrical frustration on the lattice can give rise to qualitatively new long-wavelength behavior that escapes a conventional gauge-theoretic description, without resorting to numerical simulations or uncontrolled approximations.

\para Here we consider the effect of quantum fluctuations for bosons in a honeycomb lattice under cluster-charging interaction at commensurate fractional fillings (1/3, 2/3, 1, 4/3, and 5/3) as shown in Fig. \ref{fig:model}~a) schematically. The strong coupling limit of this model, first inspired by the extended Wannier orbitals in twisted bilayer graphene\cite{zhang2022fractional}, is equivalent to the problem of densely tiling triangles on faces of a triangular lattice\cite{trimer-bethe}. Finding a tiling solution for a finite region with general tiles is a computationally hard problem. However, we can leverage the exact Bethe ansatz to predict exponential degeneracy in the tiling solutions in this case. 

\para In this paper, we study whether the degenerate configuration space of the complex optimization problem can impact the quantum ground state properties even amidst finite quantum fluctuations. We find the powerful notion of tiling invariance by Conway and Lagrias~\cite{conway1990tiling} to dictate the quantum dynamics within the constrained space of configurationsan to emergently conserve a global charge. 
This opens the possibility of a macroscopically coherent quantum ground state characterized by spontaneous breaking of an emergent $U(1)$ symmetry: the phase we term a ``supernematic" (SN). The SN exhibits a new form of intertwined order, being phase-coherent (``super-") while breaking $C_3$ rotational symmetry (``-nematic"), where the elongated shape of the resonating operator enables the latter. 
It is a rather unusual phase. It is an incompressible insulating state like a Mott insulator, hence not a superfluid. Yet it is a macroscopically phase-coherent state with gapless excitations.   

\para The rest of the manuscript is organized as follows.
In Sec.\ref{sec:model}, we introduce the cluster charging model and discuss the smallest resonating operator with an elongated, lamniscate shape (See Fig.\ref{fig:model}b). 
In Sec.\ref{sec:proof}, we prove the exactness of the emergent quantum number $\tilde{N}\equiv N_A-N_B$, the sublattice polarization, for the manifold satisfying cluster-charging constraints at a filling $2/3$ of the honeycomb lattice. We do this by mapping the configurations in the constrained Hilbert space to a tiling problem with a fixed tile set and using the Conway-Lagrias boundary invariant in Ref.\cite{conway1990tiling}. 
In Sec. \ref{sec:eff}, we derive the effective field theory that can address the quantum phase transition associated with the spontaneous symmetry breaking of the emergent $U(1)$ symmetry. This is done through mapping the microscopic problem to a three-colored double-dimer model and then using the height-field representation of the dimer model. The resulting field theory describes a problem with intertwined orders with fields transform non-trivially under the emergent $U(1)$ symmetry, the lattice translation and the $C_3$ point group symmetry. We predict a continuous transition between a columnar-ordered phase (CO) that breaks translational symmetry and $C_3$ point-group symmetry (See Fig.\ref{fig:model}c), and a new macroscopically phase-coherent phase, which breaks the emergent $U(1)$ symmetry and the $C_3$ point-group symmetry, but not the lattice translation (See Fig.\ref{fig:model}d). Hence, we refer to this new phase as supernematic (SN).
In Sec.\ref{sec:CO} we discuss the properties of the columnar ordered phase and its melting via thermal fluctuations. 
In Sec.\ref{sec:SN} we discuss the properties of the supernematic phase and its thermal melting.
The thermal phase diagram is summarized in Fig. \ref{fig:model}e). 
In Sec.\ref{sec:sum}, we summarize our results and discuss the implications.

\section{Extended Hubbard model and the cluster charging regime}
\label{sec:model}
\begin{figure*}
    \centering
    \includegraphics[width=0.9\textwidth]{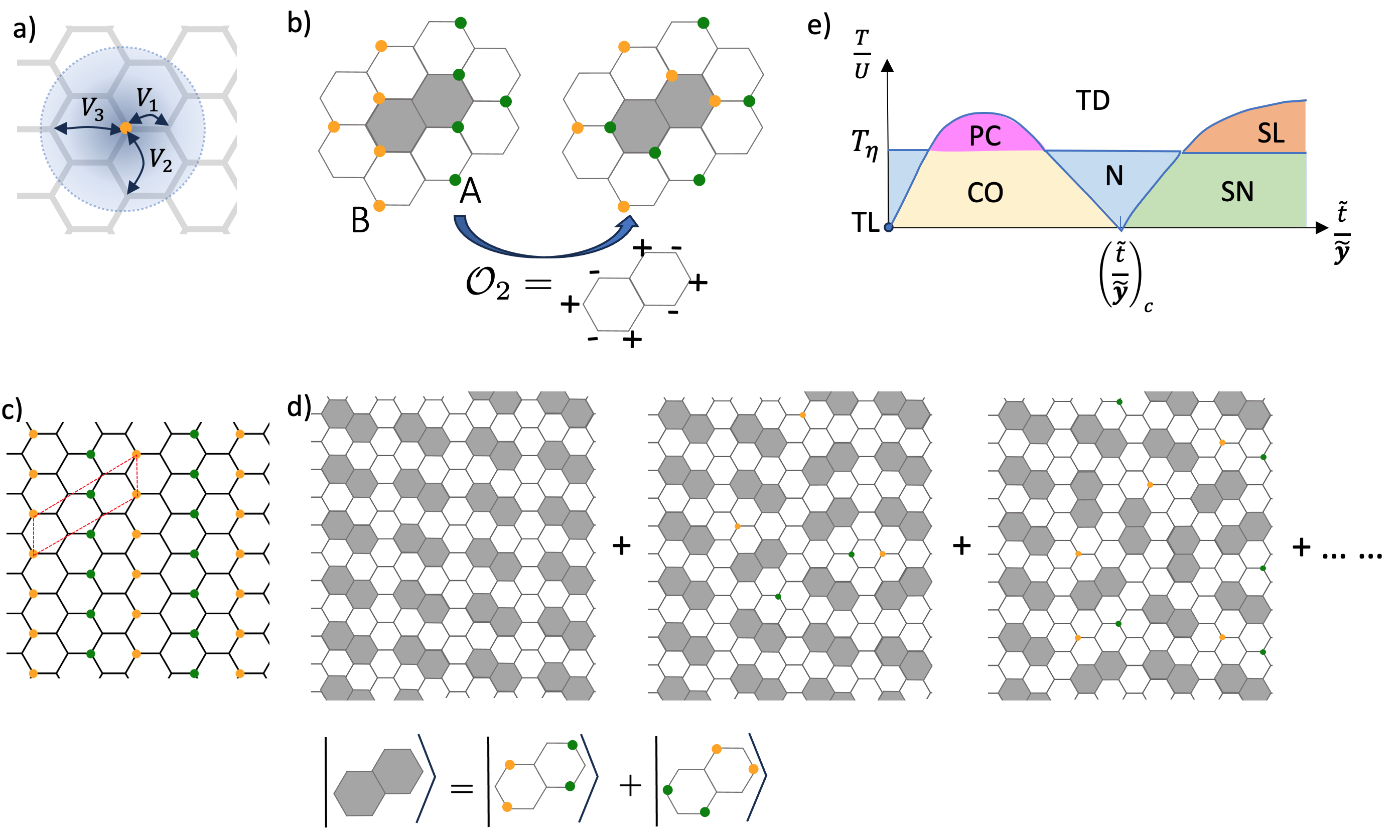}
    \caption{a) The cluster charging model, with repulsive interaction upto third nearest neighbor. The blue shaded area denotes the range of the interaction. b) The smallest local move that changes the particle configuration locally, by applying the $\mathcal{O}_2$ operator, whose $\pm$ signs denote particle creation/annihilation. c) The colomnar order with alternating colomns of occupation on A sublattices (green) and B sublattices (orange). The unitcell is indicated by the red dashed lines. d) Top: the schematic wavefunction of the resonating state, with superposition of various tilings of the local resonating patterns (grey double-plaquette). Bottom: the grey double-plaquette denotes a local resonance of the two configurations. e) Schematic phase diagram of the quantum model, where $\tilde{t}/\tilde{\mathbf{y}}$ is the tuning parameter (see Sec.\ref{sec:eff-phases}), and $T_\eta$ denotes the transition temperature of the $\vec{\eta}$-field, which is tuned by a different parameter of $\tilde{t}/\tilde{\mathbf{y}}$. Here the $T_\eta$ is picked such that all the thermal phases discussed in Sec.\ref{sec:finite_CO}  and Sec. \ref{sec:finite_SN} can be shown. For larger $T_\eta$, the area of PC and SL will shrink and these phases can disappear. TL: ``trimer liquid", which is the extensively degenerate classical ground state satisfying cluster-charging constraintes. CO: ``colomnar ordering" (panel b)). PC: plaquette Coulomb. N: Nematic phase. SL: ``Super"-fluid. SN: Supernematic (panel d)). TD: thermally disordered. 
    The quantum phase transition between CO and SN at zero temperature is proposed to be of 3D XY universality class. 
    The blue lines denote the thermal phase transition of the three-state Potts universality class. }
    \label{fig:model}
\end{figure*}

\para Consider an extended Hubbard model on the honeycomb lattice, with bosonic (or spinless fermionic) degrees of freedom on the sites. In general, the Hamiltonian for such system can be written as,
\begin{equation}
    H = \sum_{i,j} V_{\langle ij \rangle } n_i n_j + \sum_{i,j} t_{ij} b_i^\dag b_j,
\label{eq:ext_H}
\end{equation}
where $V_{\langle ij \rangle } = V_0,  V_1, V_2, ...$ denotes the interaction between particles of the same site, nearest neighbor, next nearest neighbor, etc (see Fig.\ref{fig:model} a)).  We always assume the hard-core condition, i.e. $V_0 \rightarrow \infty $.

\para The cluster-charging interaction of interest $H_U$ represents an intermediate-range interaction in the form of
\begin{equation}
    H_U = U \sum_r (\sum_{i \in \hexagon_r} n_i)^2,
    \label{eq:cluster_charging}
\end{equation}
where the $n_i$ is summed over the sites of an hexagonal plaquette labelled by $r$, and the interaction is summed over all honeycomb plaquettes. Expanding $H_U$ in terms of density-density interaction, $4 U = V_1 = 2 V_2  = 2 V_3$. 
This interaction is longer ranged than the on-site interaction of the Hubbard model, but it is not infinite-ranged like the Coulomb interaction. 
It has been shown that $H_U$ introduces non-trivial orbital geometrical frustration, resulting in extensively degenerate configurations and local $U(1) \times U(1)$ gauge symmetry \cite{trimer-bethe,Giudice2022Phys.Rev.Ba, mao2023fractionalization} in the classical strong coupling limit. This intermediate-range interaction can arise when the interaction strengths in the general Hamiltonian Eq.\ref{eq:ext_H} follow a specific hierarchy. 
For the commensurate cases with filling  $1$ and $2/3$ particles per unit cell, $H_U$ dominate when $\max\{V_1, \frac{V_2}{2}, \frac{V_3}{2}\} \gg |V_1 - \frac{V_2}{2}|, |V_1 - \frac{V_3}{2}|, |V_2 - V_3|, V_k, W$, where $k\geq 4$ denotes the further range interaction, and $W$ denotes the bandwidth. For filling $1/3$, $H_U$ dominate when $V_{1,2,3} \gg V_k, W$. 

\para In Ref.\cite{zhang2024bionicfractionalizationtrimermodel}, we showed that the classical ground state in the strong coupling limit of $H_U$  is
the so-called ``trimer liquid" state, which is an ensemble of equal probability of all the possible particle configurations that minimize $H_U$. The trimer liquid has power-law correlation functions and pinch-point structure factor of particle density. 
The question we would like to address is the fate of trimer liquid under {\it bosonic quantum fluctuations} in the cluster-charging regime. We will primarily focus on filling of $2/3$ particle per hexagonal plaquette and discuss generic commensurate fillings in Sec. \ref{sec:other_filling}. 

\para There are several differences between the cluster-charging regime of interest and two better-known strongly interacting quantum bosonic models: the Bose Hubbard model and the extended Bose Hubbard model. With the extremely short-ranged interaction of the Hubbard model, the strong coupling limit is atomic, where each site can host zero or one bosons, subject to the total fillings. Quantum fluctuation introduced through hopping drives superfluidity, resulting in the well-known phase diagram \cite{Fisher1989Phys.Rev.B} with ``Mott-lobes'' anchored by integer fillings with superfluid phase in-between, in terms of chemical potential versus hopping strength. With the further-range interaction of the extended Bose Hubbard model, quantum fluctuation usually gives rise to superfluid or supersolid phases where both break the $U(1)$ symmetry associated with the total particle number. However, such phase coherence is forbidden for the cluster-charging Hamiltonian ground state since the total particle number is related to the sum of the cluster-charging over all the plaquettes, which are local gauge symmetries. By Elitzur's theorem \cite{elitzur1975impossibility}  the usual $U(1)$ symmetry no longer can be broken spontaneously, guaranteeing a correlated insulating state for our model. Now, the question is: what else can happen to the quantum ground state of the model? 

\para The first step for considering the effect of quantum fluctuation in geometrically frustrated problems is to 
identify the minimal resonating operator that connects different configurations and the symmetry these operators may preserve. In the canonical example of the dimer model, the effect of quantum fluctuations is wildly different between the bipartite and non-bipartite lattices. 
The local resonating operators admit a $U(1)$ gauge structure for bipartite lattice while a $Z_2$ gauge structure for non-bipartite lattice \cite{SubirKagome,moessner2001short}. 
Due to the confinement of the compact $U(1)$ gauge theory in $2+1D$ \cite{POLYAKOV1977429}, the quantum dimer model on a bipartite lattice has gapped ground states that usually break lattice translational symmetry. On the other hand, 
the quantum dimer model on a non-bipartite lattice can have a gapped topologically ordered state, which is the deconfined phase of the $Z_2$ gauge theory \cite{FS79} \footnote{A recently established powerful perspective for understanding the radical difference observes 
that the local resonating operator preserves a $U(1)$ 1-form symmetry associated wth the winding number of the dimers in the bipartite lattice, while only a $Z_2$ 1-form symmetry in the non-bipartite lattice. (see e.g. Ref.\cite{higherformLSM, Pace2023Phys.Rev.B} and the reference therein) }

\para 
In the cluster-charging regime, the local resonating operators depend on the filling fraction. For the filling of $2/3$, the operator involving the smallest number of nearest neighbor hopping is the operator we denote as $\mathcal{O}_2$ in Fig.\ref{fig:model}c), because a smaller resonating operator involving only one plaquette annihilates all the states in the constraint space (see Appendix Sec.\ref{sec:app:ops} for the derivation). There are significant consequences to the extended local structure of the $\mathcal{O}_2$ taking the elongated shape (a similar scenario for filling $1/3$, which is called ``lemniscate" in Ref. \cite{mao2023fractionalization}, also discussed in Sec.\ref{sec:other_filling}). Firstly, the elongated shape means that the resonant fluctuation can occur locally in three different orientations, $\alpha=1,2,3$. 
Secondly, the resonating operator $\mathcal{O}_2$ locally shuffles the sub-lattice polarization between the two neighboring plaquettes supporting the $\mathcal{O}_2$ operator. 
Hence, in the cluster-charging regime of filling $2/3$, the effective Hamiltonian can be written as
\begin{equation} \label{eq:CR}
\mathcal{H}= - t_2 \sum_{i,\alpha} \mathcal{O}_2^\alpha(x_i) + h.c.+H_U+V_0 \sum_i n_i^2 +...,
   \end{equation}
where 
$\alpha$ denotes the different orientation of the $\mathcal{O}_2$ operator and $H_U$ refers to the cluster charging interaction in Eq.~\eqref{eq:cluster_charging}, which is a constant that can be dropped. We include the on-site Hubbard term proportional to $V_0$ to impose the hard-core condition, and the $...$ to denote higher order ring-exchange operators and other density-density interactions. 

\para We are interested in the phases of the quantum Hamiltonian $\mathcal{H}$ under finite quantum fluctuation $t_2$ and interactions.
The emergent $U(1)\times U(1)$ gauge structure we found earlier in the classical limit by mapping the classical model to coupled dimer models (see Ref. \cite{mao2023fractionalization} Appendix F, also see Sec.\ref{sec:eff})
carries over to the quantum Hamiltonian $\mathcal{H}$.
If the local $U(1)\times U(1)$ symmetry were to be the only conservation law in the problem, one would expect quantum fluctuations to drive confinement and a gapped spectrum as in the quantum dimer model on a bipartite lattice. 
However, as we will show in the next section, the $\mathcal{O}_2$ operator globally preserves an additional quantum number: the sublattice polarization, $\tilde{N}\equiv N_A-N_B$, which will be proved to remain conserved under all local terms in $\mathcal{H}$. This conservation law is independent of the 1-form symmetry of the $U(1)\times U(1)$ gauge theory, indicating a richer structure beyond conventional confinement physics in quantum dimer models. 
The conservation of $\tilde{N}$ suggests a possible spontaneous symmetry breaking phase with macroscopic phase coherence, analogous to a superfluid, coexisting with the confinement of the $U(1)\times U(1)$ gauge theory: the supernematic (SN). %We call this gapless phase supernematic (SN). 
As we show later, the SN has two features: the spontaneous symmetry breaking of $\tilde{N}$ conservation leads to gapless Goldstone mode and phase coherence, and the confinement of the $U(1) \times U(1)$ gauge field leads to the $C_3$ three-fold rotational symmetry breaking.

\begin{figure*}
    \centering
    \includegraphics[width=0.78\textwidth]{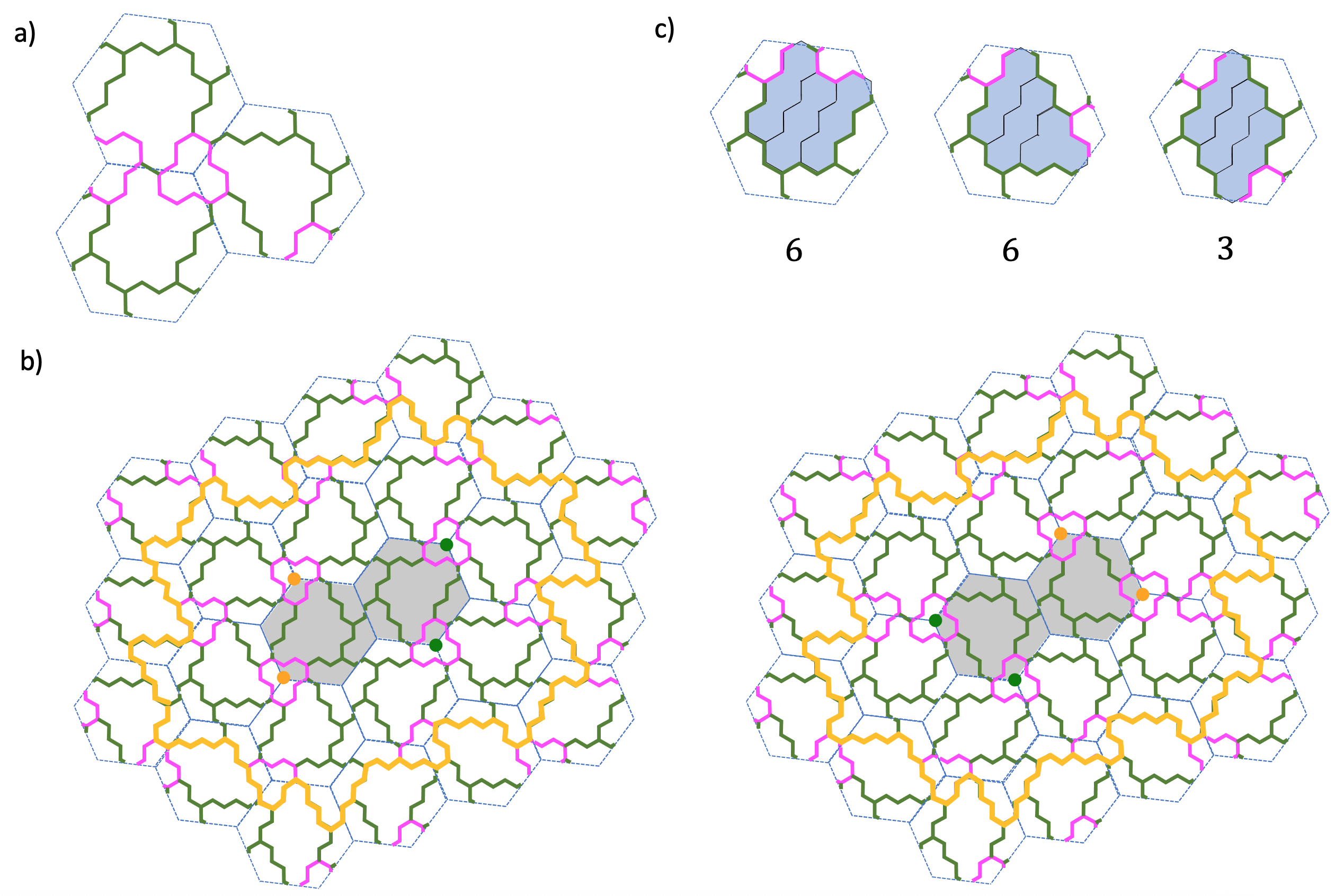}
    \caption{a) A mapping between particle configuration and tiling pattern with miniature tiles. The magenta-contoured triangular tiles maps to the sites with particle occupation. The green-contoured and mixed-color-contoured tiles map to the holes and the spaces in between particles and holes. b) The tile mapping of the configurations in a), with the same blue shaded region. The yellow line marked the boundary where the change of the configuration is contained, and the shape of the boudary remain inact under the flipping move. The grey shades and the orange and green dots mark the corresponding configuration in Fig.\ref{fig:model} b). c) The exact mapping between all the possible particle configurations of each plaquette (puzzle pieces) and the tiling pattern. There are 15 different configurations (chosing 2 out of 6). The number underneath each plaquette denotes all the confiugrations related by $C_6$ rotation. The blue shades denote the decomposition of a bigger tile into small tiles. Note that there can be multiple possible decompositions but all of them consist of tiles given in Fig.\ref{fig:tiles} a) and b).}
    \label{fig:ops}
\end{figure*}

\section{Searching for new quantum number as a tiling problem}
\label{sec:proof}
\para We seek the quantum number that is preserved under local resonating dynamics (Fig.\ref{fig:model}b) and higher order terms) by mapping the problem of allowed particle configurations to the problem of tiling. Tiling problems are a class of combinatorial problems generally considered NP-hard. The rule of the game is to cover a given bounded region with certain geometric shapes, i.e., the tiles, such that the tiles neither overlap nor have gaps in between.
 The seminal work of Conway and Lagarias \cite{conway1990tiling} shifted the focus from finding a tiling solution to establishing a no-go theorem of tiling impossibility: tiling is impossible when boundary invariants associated with the tile set are violated, where the boundary invariants are derived using the language of combinatorial group theory. 
 As we will make precise in the following, such a boundary invariant can lead to a new quantum number when dynamics is restricted to configurations satisfying cluster-charging contraints at filling $2/3$.

\para To make the mapping between tiling and particle configurations at filling $2/3$, we first consider miniature tiles that are smaller than the original honeycomb plaquettes (shapes with green and magenta contours in Fig.\ref{fig:ops}a). The intuition is to imagine "blowing up" each particle so that it acquires a geometric shape, with none of the shapes overlapping. Equivalently, we can view the tiling pattern as emerging from joining together the ``puzzle" pieces, where each piece is the physical hexagonal plaquette decorated with magenta and green lines. We can now approach the resonance of Fig.\ref{fig:model}c) as two allowed tiling configurations where only a subset of the miniature tiles are rearranged (Fig.\ref{fig:ops}b)). 

\begin{figure}
    \centering
    \includegraphics[width=0.48\textwidth]{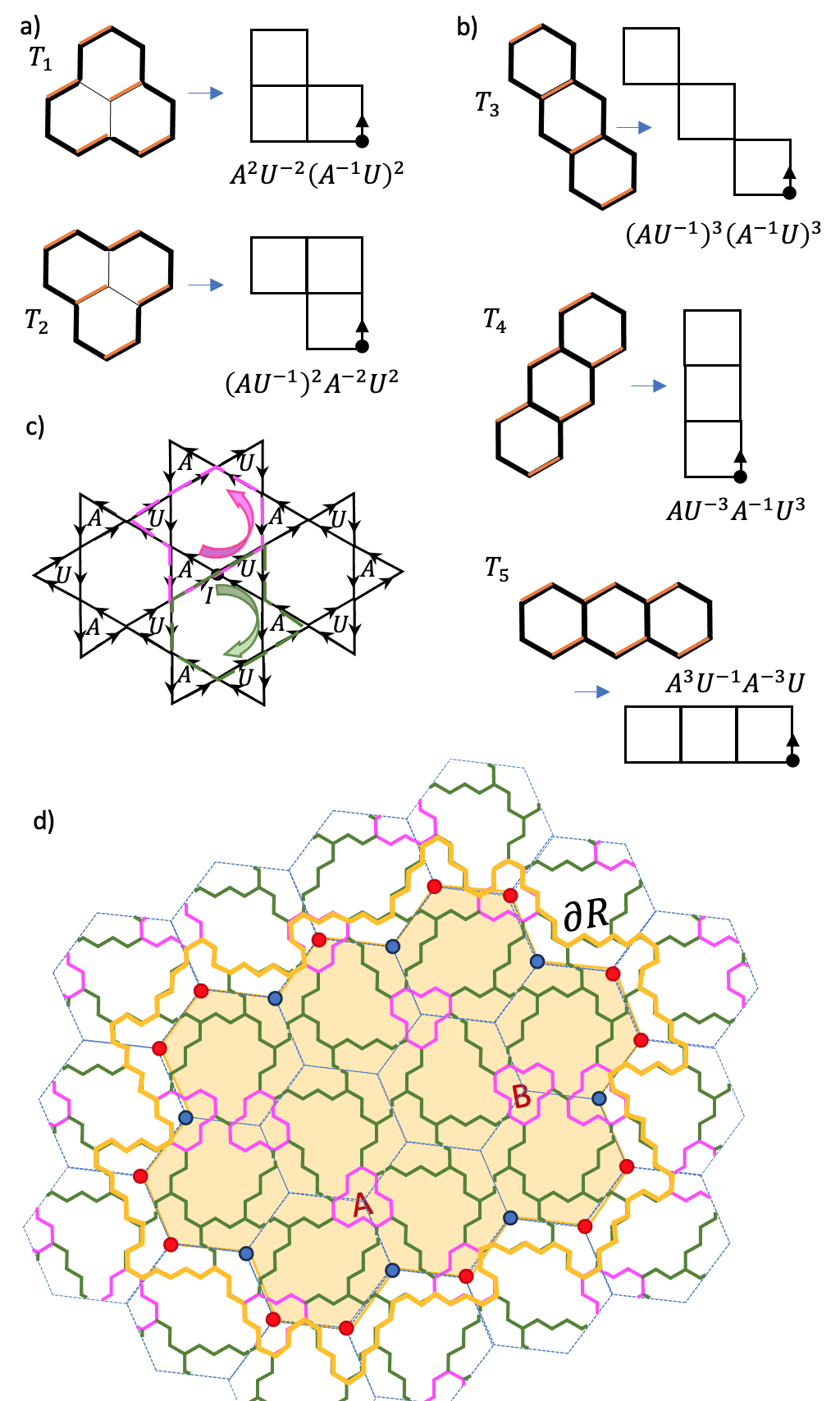}
    \caption{a) Triangular tile set $\Sigma_1$ and b) three-in-line tile set $\Sigma_2$. The black dots label the base points for the paths that generate the words written below. 
    c) Cayley graph, where the magenta and green loops correspond to the paths of $\partial T_1$ and $\partial T_2$, with the arrows indicating the direction of the paths. d) Taken from Fig.\ref{fig:ops}c. The orange line denotes the defined boundary $\partial T$ for a region that consists of orange-shaded plaquettes. The magenta tiles denote the particle occupation, with $A$ and $B$ sublattices labeled. The sites on the boundary of the orange-shaded region belong to either $S_2$ (blue dots) or $S_1$ (red dots), depending on whether they lie in two or one hexagonal plaquettes of the orange region of interest.}
    \label{fig:tiles}
\end{figure}

\para Now, to further connect to the original problem considered in Ref.\cite{conway1990tiling}, we view the bigger tiles with the green/magenta contours in Fig.\ref{fig:ops} as consisting of the smaller three-in-line tiles and triangular tiles of three cojoining plaquettes (blue shades in \ref{fig:ops}c). All the allowed particle configurations can therefore be mapped to tiling patterns by connecting the hexagonal pieces in Fig.\ref{fig:ops}c) in a boundary-matching way. The sites with particles are mapped to magenta tiles. Note that all the tiles fall into the tile sets in Fig.\ref{fig:tiles}a) and b).

\para For these tiles, the following theorem proved in Ref.\cite{conway1990tiling} connects allowed tiling configurations to a boundary invariant. 
\begin{theorem}[Conway-Lagarias]
\label{thm:CL}
    For any simply connected region $R$, if $R$ is tilable by $\Sigma_1$ (Fig.\ref{fig:tiles} a) and $\Sigma_2$ (Fig.\ref{fig:tiles} b), we can define a boundary invariant $\phi([\partial R])$. On the one hand, $\phi([\partial R])$ only depends on the boundary of $R$, denoted as $\partial R$. On the other hand, $\phi([\partial R]) = \sum_{i \in R} \phi([\partial T_i])$, where $T_i$ is the tile contained in $R$, and $\phi([\partial T_1]) = 1$, $\phi([\partial T_2]) = -1$, and $\phi([\partial T_{3,4,5}]) = 0$.
\end{theorem}
Here, tilability means there are no gaps or overlaps between the tiles, which corresponds to satisfying the cluster charging constraints while fixing the filling to $2/3$ particles per unit cell. 
In the remainder of this section, we first give a pedagogical proof of Thm.\ref{thm:CL} for completeness and then show that the tiling invariance rigorously establishes a new emergent quantum number for the quantum cluster-charging Hamiltonian of Eq.~\eqref{eq:CR}.  

\subsection{Conway-Lagarias Invariant}
\para
To construct the boundary invariant in Thm.\ref{thm:CL},  Ref.\cite{conway1990tiling} maps the boundary of the tiling to another ``space", where the boundaries of the different tiling types carry different winding numbers. 

\para First, we need an algebraic description of the tiles and the tiling problem. The relevant tile set for $2/3$-filling (Fig.\ref{fig:ops})
can be classified into two sets, triangular set $\Sigma_1$ (Fig.\ref{fig:tiles}a) and three-in-line set $\Sigma_2$ (Fig.\ref{fig:tiles}b). It turns out to be convenient to view the tiles $T_i$ as tiles on the square lattice by ``shrinking" along one orientation of the bonds on the honeycomb lattice, as illustrated in Fig.\ref{fig:tiles}a and b. We glue the two sites connected by the orange bonds on the honeycomb lattice and reshape the remaining edges to be horizontal and vertical to obtain the corresponding tiles on the square lattice, which are called ``polyominoes". 

\para Then, we need to characterize different ``shapes" of the tiles. One way to achieve this is to describe the paths along the boundary of the tiles by the words (elements) in free group. On the square lattice, we consider a free group $F = \langle A, U\rangle$, where $A$ and $U$ are the two generators, denoting going right and going up. Next, we fix a base point $\vec{e}_i$ on the boundary of a tile $T_i$, and denote the ``word" generated by going along the boundary counterclockwise as $\partial T_i(\vec{e}_i)$, with the corresponding generators multiplied to the left. We call this {\it oriented boundary} of $T_i$. For example, the oriented boundary of $T_1$ can be written as $\partial T_1(\vec{e}_1) = A^2 U^{-2} (A^{-1} U)^2$ (see Fig.\ref{fig:tiles}a and b for other $\partial T_i$'s). It is readily seen that if we choose a different base point on the lattice, say $\vec{e}'_i$, the oriented boundary $\partial T_i(\vec{e}_i')$ would be the conjugate of the previous word, and can be written as $W^{-1} (\partial T_i(\vec{e}_i)) W$, where $W\in F$. Hence, for any simply connected region $R_s$, it is convenient to define the {\it combinatorial boundary} of $R_s$ as the conjugacy class in $F$ that contains all the oriented boundaries $\partial R_s (\vec{e})$'s, that is $[\partial R_s] = \left\{W^{-1} \partial R_s(\vec{e}) W, W\in F\right\}$.

\para  Now we join the tiles to form a tiling pattern. First if one joins two tiles $T_i$ and $T_j$, the boundary word of the joined region can always be written as $W_i^{-1} (\partial T_i(\vec{e}_i)) W_i W_j^{-1} (\partial T_j(\vec{e}_j)) W_j $, with $W_{i,j}\in F$. We can then proceed to put more tiles together. Eventually, we observe that for any simply connected tilable region ${R}$, its boundary word can always be decomposed into boundary words of the single tiles inside ${R}$, that is, 
\begin{equation}
\partial {R} = \prod_{i \in {R}} W_i^{-1} (\partial T_i(\vec{e}_i)) W_i.
\label{eq:boundary_path}
\end{equation}
In the above equation, the l.h.s. depends only on the path along the boundary $\partial {R}$, whereas the r.h.s. depends on the $\partial T_i$'s, which in turn depend on the tiles inside the region ${R}$. The $W_i$'s contain information of how these tiles glue together. Hence, we can define {\it tile group} $T(\Sigma)$ as
\begin{equation}
    T(\Sigma) = N(\langle \partial T_i(\vec{e}_i)\rangle),
\end{equation}
where $N(\dots)$ denotes the normal subgroup in $F$. From our previous argument, if ${R}$ is tilable by $\Sigma$, $\partial {R} \in T(\Sigma)$.

\para Next, in order to find the so-called boundary invariant, we only need to find a homomorphism between the tile group and integers. We achieve this by first finding a larger group $H$ that contains the tile group $T(\Sigma)$.

\begin{equation}
    H = N(\langle A^3, U^3, (U^{-1} A)^3\rangle).
\end{equation}
The presentation of $H$ is less important to us since we will work with the Cayley graph $\mathcal{G}(F/H)$ in the following.
The Cayley graph $\mathcal{G}(G)$ is a graph with directed edges for visualizing the structure of the group $G$. We define the identity element $I$ as the origin of the graph. For each vertex, the outgoing edges denote the generators of the group and the ingoing edges denote their inverses. The Cayley graph $\mathcal{G}(F/H)$ is shown in Fig.\ref{fig:tiles}c, which is a Kagome lattice. The edges are labeled with the generators $A$ and $U$. Let us consider a directed path on the graph and define the following rule: If the path's direction matches the edge it crosses, the corresponding generator is multiplied; otherwise, the inverse of the generator is multiplied. Following this rule, we can map elements in $T(\Sigma)$ to paths on the Cayley graph in Fig.\ref{fig:tiles}c. Since $T(\Sigma) \subset H$, these paths are closed loops, as the magenta and green loops shown in Fig.\ref{fig:tiles}c.

\para For a directed close path $L$ on the Cayley graph, we can always define the winding number  $\phi(L) $ to be the number of counterclockwise loops around the hexogonal cells enclosed by the closed directed path $L$. In particular, we find that for the combinatorial boundaries of the tiles, $\phi([\partial T_1])  = 1$, $\phi([\partial T_2]) = -1$, and $\phi([\partial T_{3,4,5}])=0$. It is straightforward to show that $\phi$ is a homomorphism by showing $\phi(AB) = \phi(A)+ \phi(B)$, for $A,B \in T(\Sigma)$. Combining the definition of $\phi$ and Eq.\ref{eq:boundary_path}, it is readily seen that for any tilable region $R$, 
\begin{equation}
    \phi([\partial R]) = \sum_{i\in R} \phi([\partial T_i])~\text{(Thm.\ref{thm:CL})}.
\end{equation}

\subsection{Quantum number for filling $2/3$}
\para Thm.\ref{thm:CL} by itself is a statement regarding each individual tiling patterns, which we have mapped to the particle configuration at filling $2/3$.
In the following, we will show that Thm.\ref{thm:CL} indeed gives rise to a conservation of the total sublattice particle number difference, $N_A - N_B$, which should be thought of as a good quantum number of a generic local Hamiltonian acting within the constrained Hilbert space. The same conclusion also holds for filling $1/3$ and the proof is given in App.\ref{sec:one-third}.

\para
To connect the winding number $\phi([\partial R])$ in Thm.\ref{thm:CL} to physical quantities, consider an arbitrary simply connected region, for example, the orange shaded region in Fig.\ref{fig:tiles}d). 
Any particle configuration in the classical ground state of $H_U$ at the fixed filling maps to a proper tiling of the region with the tile set $\{T_1,\cdots,T_5\}$ shown in Fig.\ref{fig:tiles}a)-b)
where the particles map to $T_1$ and $T_2$ marked in magenta, and the space between particles maps to regions with green/magenta contours which can be built with the full tile set $\{T_1,\cdots,T_5\}$ as shown in Fig. \ref{fig:ops}c). To apply Thm.\ref{thm:CL}, we need to define the boundary of the region carefully.
The tiling boundary $\partial R$ (yellow line) is naturally defined as the outer contour of the magenta and green tiles covering the physical lattice sites along the boundary of the region of interest. Some sites touch only one physical plaquette in the region of interest. These sites form the set $S_1$ and are marked in red. Some sites touch two physical plaquettes in the region of interest. These sites form the set $S_2$ and are marked in blue.  
Now we can apply
the Thm.\ref{thm:CL} to the tiling boundary $\partial R$, to find the CL tiling invariance for the tile set $\{T_1,\cdots, T_5\}$: 
\begin{equation}
\phi([\partial R]) = (N_{T_1} - N_{T_2})_R,
\label{eq:inv}
\end{equation}
where the right hand side is the number of $T_1$ tiles minus the number of the $T_2$ tiles within the region $R$. The simplicity emerges from the winding number of the tiles $T_{3,4,5}$ being zero, i.e., $\phi([\partial T_{3,4,5}])=0$. 

\begin{figure}
    \centering
    \includegraphics[width=0.8\linewidth]{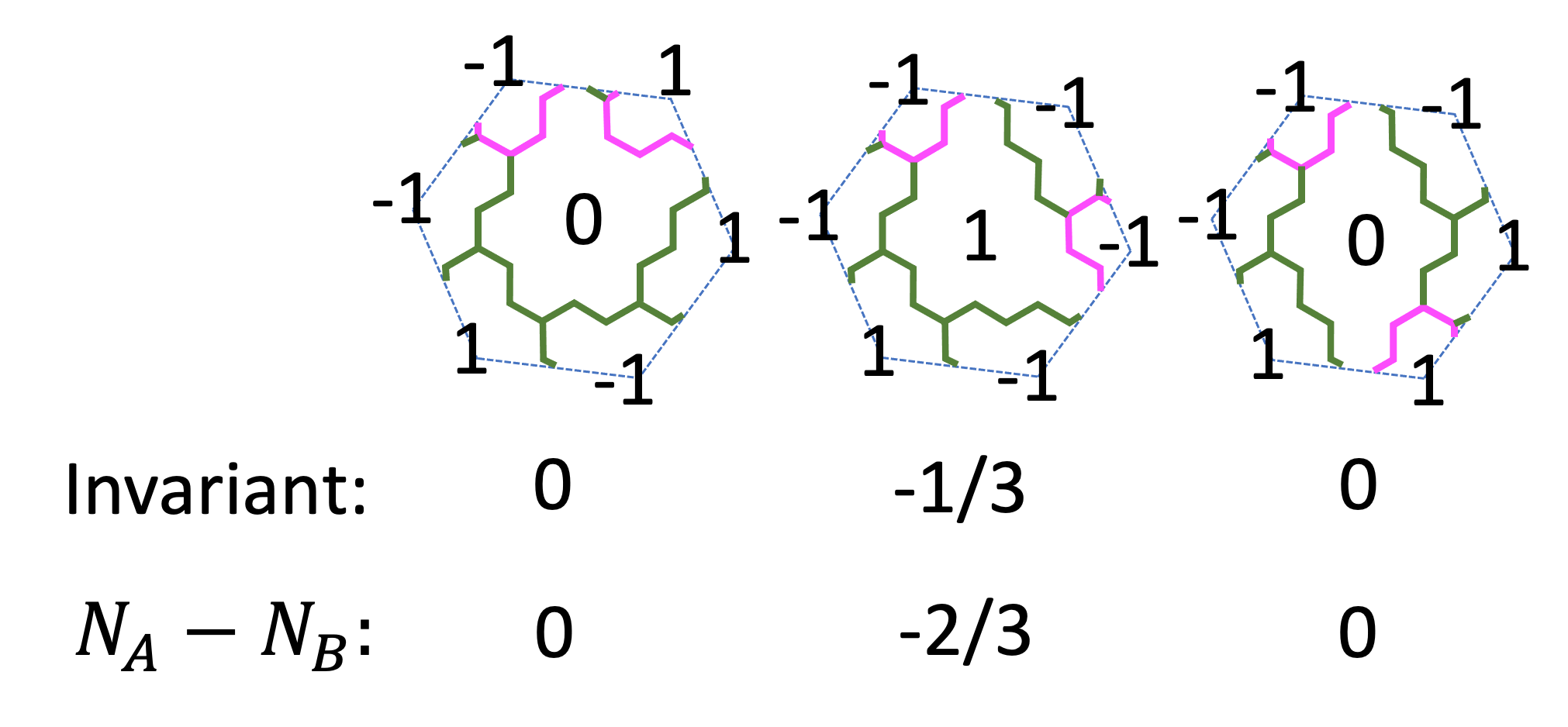}
    \caption{Assignment of the ``puzzle" invariant $\Phi_p$ for three configurations. The integers denote the boundary invariant of the corresponding regions. The invariant of the puzzle piece is calculated by summing the middle number and $1/3$ of the numbers at the boundary. The ``$N_A - N_B$" row denotes $1/3$ of the total particle occupation $N_A - N_B$ of the above hexagon, magenta tiles being the locations of the physical particles. Here only the puzzle with $N_A = 0, N_B = 2$ is shown, the piece with $N_A = 2, N_B = 0$ can be obtained by a $60^\circ$ rotation of the former one, giving $\Phi_p = 1/3$. }
    \label{fig:puzzle_inv}
\end{figure}

\para To use the CL invariance Eq.~\eqref{eq:inv} to derive a global quantum number for the Hamiltonian $\mathcal{H}$ at filling $2/3$, we need to deal with the complexity arising from the ``puzzle'' pieces being composites including $T_1$ and $T_2$ tiles. While $T_1$ and $T_2$ tiles marked in magenta in Fig.~\ref{fig:tiles}d) correspond to $A$ and $B$ sublattice site occupation, respectively, $T_1$ and $T_2$ inside the green tiles do not correspond to physical particles. To carefully deal with the complexity,
%Next, we relate $N_{T_1} - N_{T_2}$ to the physical quantity $N_A - N_B$ within the boundary $\partial R$. 
%To this end,
we assign a ``puzzle'' invariant $\Phi_p$ to each of the fifteen puzzle pieces by summing two contributions: 
(i) the boundary invariant in the middle region and 
(ii) one-third of the boundary invariants of the tiles surrounding the vertices of the hexagon (see Fig.~\ref{fig:puzzle_inv}). 
Here, we use the CL invariance Eq.~\eqref{eq:inv} to calculate the boundary invariant of each of the tiles. 
As shown in Fig.~\ref{fig:puzzle_inv}, each particle occupying the A (B) sublattice contributes $+1/6$ ($-1/6$) to $\Phi_p$. 

\para 
Now we can use Thm.~\ref{thm:CL} to relate the boundary invariant $\partial R$ to a sum of puzzle invariants for the puzzles enclosed in the region $R$ and the boundary tile contributions as
%By definition, summing over the puzzle invariants within region $R$ is also related to the boundary invariant of $\partial R$,
\begin{equation}
   \phi([\partial R]) 
   =  \sum_{i \in R} \Phi_p(i) 
   +  \frac{2}{3} \sum_{i \in S_1} \phi_{i} 
   +  \frac{1}{3} \sum_{i \in S_2} \phi_{i},
   \label{eq:phiR_vs_phiP}
\end{equation}
where $\phi_i$ denotes the winding number of the tiles at the site $i$ on the boundary. Next, we observe the relationship between the puzzle invariants and the sublattice polarization $\tilde{N}\equiv N_A-N_B$.
While $\frac{1}{3}\tilde{N}$ can be $\{-2/3, 0, 2/3\}$ for each plaquette $p$, corresponding $\Phi_p$ are $\{-1/3, 0,1/3\}$ as depicted in Fig.~\ref{fig:puzzle_inv}. Hence summing over $\Phi_p$ within region $R$ contains summing over $\frac{1}{2}\tilde{N}$ in the interior since each vertex is touched upon three times by the hexagonal plaquettes.
At the boundary, however, how many internal plaquettes in the region $R$ share the particles (the magenta tiles) depends on which set the boundary site $i$ belongs to. When the site $i$ is in $S_1$ (red), the magenta tile is touched by only one internal plaquette, and the counting from summing over $\Phi_p$ misses a factor of $\frac{1}{3}$. On the other hand, when the site $i$ is in $S_2$ (blue), the magenta tile is touched by two internal plaquettes, and the missing factor is $\frac{1}{6}$. Hence we arrive at the full counting of the sublattice polarization of the region $R$ (including both the interior and the boundary) as
\begin{equation}
    \frac{1}{2}(N_A - N_B)_R 
    = \sum_{i \in R} \Phi_p(i) 
    + \frac{1}{3} \sum_{i \in S_1} \phi_{i,M} 
    + \frac{1}{6} \sum_{i \in S_2} \phi_{i,M},
    \label{eq:Ntilde_vs_phiP}
\end{equation}
where $\phi_{i,M}$ denotes the winding number of the magenta tiles enclosing site $i$ on the boundary.

\para Now we can eliminate the $\Phi_p$ terms between Eqs.~\ref{eq:phiR_vs_phiP} and \ref{eq:Ntilde_vs_phiP} to obtain the first major result of this paper:
\begin{equation}
\begin{split}
 &\frac12 (N_A- N_B)_{R} \\
 =& \phi([\partial R]) -\frac23 \sum_{i\in S_1} (\phi_i - \frac12 \phi_{i, M})- \frac13 \sum_{i \in S_2} (\phi_i- \frac12 \phi_{i,M} ).
\end{split}
\label{eq:b_inv_23}
\end{equation} 
The left-hand side of Eq.~\ref{eq:b_inv_23} is a purely ``bulk'' physical quantity determined by the configuration within region $R$, while the terms on the right-hand side are boundary contributions. In the sense that there is a bulk charge that can be related to the boundary properties, the relationship is reminiscent of topological invariants. However, unlike topological invariants with underlying gauge theory, the $\tilde{N}_R \equiv (N_A - N_B)_R$ depends on the shape of the boundary. Instead, Eq.~\ref{eq:b_inv_23} implies a conserved global quantum number associated with the total $\tilde{N} = N_A - N_B$. 
To see this, consider any local operator $\mathcal{O}$ with finite support in an infinite system. 
We can always choose a region $R$ that fully encloses the support of $\mathcal{O}$ such that its action leaves the boundary configuration unchanged. 
Consequently, acting with $\mathcal{O}$ does not modify the right-hand side of Eq.~\ref{eq:b_inv_23}, implying that the sublattice particle number within $R$ is conserved from the left-hand side. 
Since $\mathcal{O}$ acts only inside $R$, the configuration outside $R$ remains unaffected, and thus the total $\tilde{N}$ in the system is conserved. 
Although the precise support of $\mathcal{O}$ and the corresponding choice of $R$ may vary, as long as $\mathcal{O}$ remains local (its support does not scale with system size), one can always identify a suitable region $R$ satisfying the above condition. Since $\mathcal{O}$ redistributes $\tilde{N}$ locally, only the total $\tilde{N}$ is conserved, making $\tilde{N}$ a distinct global symmetry rather than the local conservation following from the cluster-charging constraints. Hence, for any effective Hamiltonian $\mathcal{H}$ acting on the constrained Hilbert space with local terms, the global quantity $\tilde{N}$ is a good quantum number.

\section{Three-colored double-dimer model}
\label{sec:eff}
\para Previously we showed that, at filling $2/3$ within the constrained Hilbert space, the sublattice imbalance $\tilde{N} \equiv N_A-N_B$ emerges as a conserved $U(1)$ charge. Taking into account this emergent symmetry with the lattice symmetry allows us to derive an effective field theory, starting from the mapping of the quantum Hamiltonian to a three-colored double-dimer model. The resulting field theory has two phases: the columnar ordering phase (CO) and the supernematic phase (SN). 

\subsection{Microscopic mapping between particle occupation and dimers}
\para The cluster-charging constraints can be recast into the more familiar constraints in dimer models by viewing the particle as three conjoined dimers. We used this mapping in Ref.\cite{zhang2024bionicfractionalizationtrimermodel} to derive the 2D statistical field theory for the classical cluster-charging model at filling $1/3$, where only two height fields are independent at long wavelength since the fluctuations around the constraint relating the three height fields are irrelevant in the classical theory. We now generalize this mapping to filling $2/3$ and consider the effect of quantum fluctuations. Quantum theory lives in $2+1$D, and we will find three independent height fields because the quantum fluctuation of the constraint is gapless.

\begin{figure}
    \centering
    \includegraphics[width=0.95\linewidth]{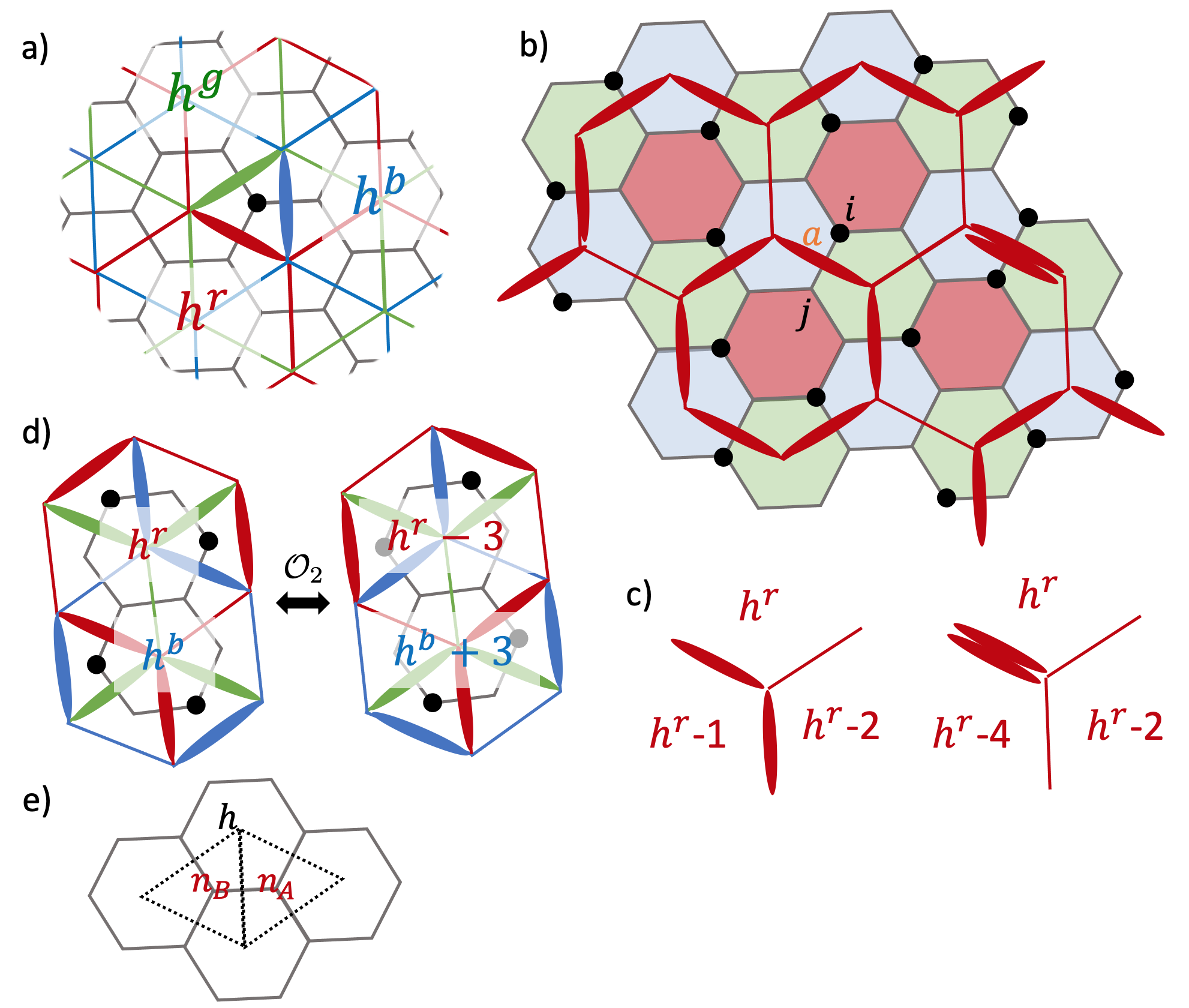}
    \caption{Mapping between trimer model at $2/3$-filling to the three-colored double-dimer model. a) Mapping between one particle and corresponding three dimers, with the height field defined at the plaquette centers of the corresponding colored hoenycomb lattices. b) An example of particle configuration and the corresponding double-dimer covering (only red is shown). The three-coloring of the original hexagonal plaquettes denotes support for the height field of the corresponding color. c) Height field assignment of the double-dimer model. d) Height change induced by applying the local move operator $\mathcal{O}_2$. e) Particle numbers on different sublattices ($n_A$ and $n_B$) are related to the sum of the three height fields $h$ (of different colors) around the sites indicated by the dashed lines. }
    \label{fig:double-dimer}
\end{figure}

\para First, we view the dual lattice of the physical honeycomb lattice as a set of three interpenetrating honeycomb lattices, labelled as red, green, and blue in Fig.\ref{fig:double-dimer} a). We then map the particle occupation to three dimer occupations (red, blue and green dimers in Fig.\ref{fig:double-dimer} a)). The mapping between the dimer occupation and physical particle number can be written as $d_a^r = n_i +n_j$, where $d_a^r$ denotes the ``red" dimer occupation number at bond $a$ and $n_{i,j}$ are the particle numbers at site $i,j$, which are the sites linked by the edge of the physical lattice intersected with bond $a$, as noted in Fig.\ref{fig:double-dimer} b). The analogous expressions also hold for the green and blue lattices. 
At filling $2/3$, the cluster-charging constraints impose ``two-dimers per site" conditions on the colores honeycomb lattice with double-occupation of dimers allowed (illustrated in Fig.\ref{fig:double-dimer} b) for the red lattice). The corresponding dimer model of each colored honeycomb lattice is the so-called ``double-dimer" model, because the configuration can be thought of as overlaying two dimer coverings with one dimer touching each vertex \cite{kenyon2014conformal}.

\para Now we introduce the height fields. In the dimer model, height fields are integer-valued fields defined at the centers of the plaquettes to encode the dimer constraints \cite{blote1982roughening, nienhuis1984triangular}. Since the double-dimer configuration can be decomposed into two single-dimer ones, Ref.~\cite{WilkinsPhysRevB.102.174431} introduces two height fields for each of the replicas. Here, for the cluster-charging model, only the sum of the two heights is relevant since no physical operators can distinguish the two replicas. We will call this summation ``the height field" in the following. The height fields of the red double-dimer model are defined at the centers of the hexagonal plaquettes, with the following rules (Fig.\ref{fig:double-dimer} c)): when going around the vertex emitting ``Y" shaped edges clockwise, if no dimer/one dimer/two dimers is crossed, height is changed by $-2$/$+1$/$+4$. The height assignment is the same for going around another type of sublattice counterclockwise. Under such rules, the action of the $\mathcal{O}_2$ operator is to shift the height field of one color by $-3$ and the height field of another color by $+3$ (Fig.\ref{fig:double-dimer} d)). 
Now the total $\tilde{N}\equiv N_A - N_B$ conservation proved in section\ref{sec:proof} maps to the global conservation of the total height fields $h_r + h_g + h_b$. 

\para We now express the Hamiltonian of our cluster charging model Eq.\ref{eq:CR}, in terms of the heights $h_{l'}^{r}$ and its canonical conjugate $\theta_{l}^r$ with 
$[\theta_{l}^{r}, h_{l'}^{r}] = i \delta_{ll'}$, where $l,l'$ label the sites and $r$ labels the color.
The hopping-induced quantum fluctuation associated with the resonance operator $\mathcal{O}_2$ can be 
 written as,
\begin{equation}
    H_t = - t_2 \sum_{l,\alpha} \cos{\left(3 \theta_{l}^r - 3 \theta_{l+\alpha}^g\right)} + (\text{permutation of colors}),
    \label{eq:H_t}
\end{equation}
where $l, l+\alpha$ label the center of the hexagonal plaquette at position $l$ and its neighbor. Since different colors reside in different sublattices, we will drop the color label for simplicity. For visual purpose, sometimes we color the hexagonal plaquettes to remind ourselves of the coloring of the height fields defined at center of the plaquettes (Fig.\ref{fig:double-dimer} b)).

\para To map the density-density interaction to heights, we first express the particle occupation in terms of the height fields (upto a constant shift of the height field of the same color, which we have fixed),
\begin{equation}
    n_A = - \frac{1}{3}\sum_{\triangleright} h + \frac{1}{3},~ n_B =  \frac{1}{3}\sum_{\triangleleft} h + \frac{1}{3}, \label{eq:p_duality}
\end{equation}
where the $\triangleright/\triangleleft$ is the shorthand notation for summing the three height fields surrounding the site of the corresponding particle (see Fig. \ref{fig:double-dimer} e)), where the color labels are implicit. The constant $\frac{1}{3}$ takes care of the filling fraction. The representation of the particle number using the fields on dual lattice can also be viewed as a plaquette duality, utilized for studying various spin models with ring exchange interactions \cite{Paramekanti2002Phys.Rev.B, Myerson-Jain2022Phys.Rev.Lett.a}. The additional link to the height fields is originated from the cluster-charging constraints.
Now, the on-site Hubbard term $H_0$ can be written as,
\begin{equation}
    H_0 = V_0 \left[ \sum_i \left(\sum_{\triangleright_i} h \right)^2 + \sum_j \left(\sum_{\triangleleft_j} h \right)^2 \right],
\end{equation}
where $i,j$ denote the $A,B$ sublattice sites and $\triangleright_i,\triangleleft_j$ denote the three plaquettes surrounding these site, and the constant shift in Eq. \ref{eq:p_duality} for A and B sublattices gives rise to a constant term that we drop. The density-density interaction beyond on-site terms can be expressed in a similar way.

\subsection{Coarse-grained height fields and their properties under lattice symmetries}
\para Now, let us focus on the limit where $t_2$ and $V_0$ are large. $H_t$ favors configurations with a uniform $\theta$-field for all the colors, while $H_0$ favors an alternating pattern of the height fields of three colors, due to the three-sublattice structure in the relationship between $h$ and $n$. Therefore, the long wavelength behavior needs to capture fluctuations near both the $\Gamma$-point ($k=0$) and the $K$-point ($k = (0, \frac{4\pi}{3})$) of the Brillouin zone: 
\begin{equation}
\begin{split}
 \theta_l &\sim \theta_0(x_l) + \vartheta(x_l) e^{i K \cdot x_l} + \bar{\vartheta}(x_l) e^{-i K \cdot x_l}\\
 h_l &\sim h_0(x_l) + \eta(x_l) e^{i K \cdot x_l} + \bar{\eta}(x_l) e^{-i K \cdot x_l},
\end{split}
\end{equation}
with the conjugacy pairs $(\theta_0, h_0)$ and $(\vartheta, \eta)$. Here $h_0$ is a real scalar field and $\eta$ is a complex scalar field.

\para The $h_0$ and $\eta$ fields are nothing but a parametrization of the $h^{r,g,b}$ fields. Namely,
\begin{equation}
\begin{split}
        h^r_l &\sim h_0(x_l)  + \eta(x_l) + \bar{\eta}(x_l)\\
        h^g_l &\sim h_0(x_l)  + \eta(x_l)e^{-i \frac{4\pi}{3}} + \bar{\eta}(x_l) e^{i \frac{4\pi}{3}}\\
        h^b_l &\sim h_0(x_l) + \eta(x_l) e^{i \frac{4\pi}{3}} + \bar{\eta}(x_l) e^{-i \frac{4\pi}{3}},
\end{split}
\label{eq:rgb_0eta}
\end{equation}
similarly for the relationship between $\theta^{r,g,b}$ and $\theta_0$, $\vartheta$ and $\bar{\vartheta}$. Here we view $h_0$ and $\eta$ as coarse-grained height fields. Due to the dimer constraints, the microscopic heights always fluctuate at the length scale of the lattice constant of the colored honeycomb lattice. Conceptually, the coarse-grained height fields $h_0$ and $\eta$ are obtained by smearing the fluctuations due to the dimer constraints, but the procedure is not relevant for us at this point. What we are interested in are the forms of physical operators in terms of the coarse-grained fields and their properties under lattice symmetries.

\para First, the compactification radius of $h^{r,g,b}$ is $3$ for our double-dimer model, since the uniform shift by three gives the same dimer configuration (same as the dimer model on honeycomb lattice \cite{Fradkin2004Phys.Rev.Bb}). Therefore, the physical operators can only be $\partial h^{r,g,b}$ (and higher order derivatives), the vertex operators $\mathcal{V}_{\vec{m}}(x) = e^{i \frac{2\pi}{3} \vec{m} \cdot \vec{h}(x)}$, where $\vec{m} = (m^r,m^g, m^b)$ is a vector of integers and $\vec{h} = (h^r, h^g, h^b)$, and their combinations.

\para The transformation of the coarse-grained fields under lattice symmetries can be derived from the ``flat" configurations of $h^{r,g,b}$'s. These are columnar orders of the double-dimer model. However, not all columnar order combinations are physical, since not all double-dimer configurations can be mapped to physical ones. There are nine physical columnar orders in total, related to the configuration in Fig.\ref{fig:model} b) by lattice translations and rotations. The symmetry transformation of the coarse-grained height fields need to account for not only the symmetry properties of the microscopic heights, but also the transformation of the average height, analogous to the discussion of the dimer model \cite{Fradkin2004Phys.Rev.Bb, fradkin2013field}. The detail is discussed in the Appendix Sec.\ref{sec:app:sym}. We will quote the results below. Under lattice symmetry,
\begin{equation}
    \begin{split}
        &\mathcal{T}_1:~h_0(x) \rightarrow h_0(x) - 1;~ \eta(x) \rightarrow \eta(x) e^{-i \frac{4\pi}{3}}\\
        &\mathcal{T}_2:~h_0(x) \rightarrow h_0(x) - 1;~ \eta(x) \rightarrow \eta(x) e^{i \frac{4\pi}{3}}\\
        &\mathcal{C}_3:~h_0(x) \rightarrow h_0(C_3 x); ~ \eta(x) \rightarrow \eta(C_3 x) - i \frac{\sqrt{3}}{3}\\
        &\mathcal{I}:~ h_0(x) \rightarrow - h_0(-x);~ \eta(x) \rightarrow - \bar{\eta}(-x),
    \end{split}
    \label{eq:trans:h}
\end{equation}
where $\mathcal{T}_1$ and $\mathcal{T}_2$ are translation along $\vec{a}_1 = (\frac{\sqrt{3}}{2}, \frac{1}{2})$ and $\vec{a}_2 = (0, 1)$, $\mathcal{C}_3$ is the counterclockwise rotation by $120^\circ$, and $\mathcal{I}$ is the inversion, both around center of the red plaquette. All the other lattice symmetries can be viewed as the combination of the above transformations. The transformation of $h_0$ and $\eta$ are subject to $h^{r,g,b} \sim h^{r,g,b} + 3 $, respectively, which we omit in the above to simplify the bookkeeping.

\para Hence, the vertex operators $\mathcal{V}_{\vec{m}}$ transforms under the following:
\begin{equation}
    \begin{split}
        &\mathcal{T}_1:~ \mathcal{V}_{(m^r, m^g, m^b)}(x) \rightarrow e^{- i \frac{2\pi}{3} (m^r + m^g + m^b)}\mathcal{V}_{(m^b, m^r, m^g)}(x)\\
        &\mathcal{T}_2:~ \mathcal{V}_{(m^r, m^g, m^b)}(x) \rightarrow e^{- i \frac{2\pi}{3} (m^r + m^g + m^b)}\mathcal{V}_{(m^g, m^b, m^r)}(x)\\
        &\mathcal{C}_3:~ \mathcal{V}_{(m^r, m^g, m^b)}(x) \rightarrow e^{- i \frac{2\pi}{3} (m^g - m^b)}\mathcal{V}_{(m^r, m^g, m^b)}(C_3 x)\\
        &\mathcal{I}:~ \mathcal{V}_{(m^r, m^g, m^b)}(x) \rightarrow \mathcal{V}^\dag_{(m^r, m^b, m^g)}(-x).
    \end{split}
    \label{eq:V_sym}
\end{equation}

\subsection{Effective field theory description}
\label{sec:eff-phases}
\para Taking into account the lattice symmetries, the effective $2+1D$ Lagragian at zero temperature can be written as,
\begin{equation}
\mathcal{L}[\theta_0, \vec{\vartheta}, h_0,\vec{\eta}] = \mathcal{L}_1[\theta_0, h_0] + \mathcal{L}_2[\vec{\vartheta}, \vec{\eta}]+ \mathcal{L}_3[h_0,\vec{\eta}]+...,
\label{eq:eff_L}
\end{equation}
where
\begin{equation}
\begin{split}
    &\mathcal{L}_1[\theta_0, h_0] = i \theta_0 \partial_\tau h_0 + \tilde{t} (\partial_\mu \theta_0)^2 + \tilde{v}_0(\partial_\mu h_0)^2 - \tilde{y}_0 \cos{(2 \pi h_0)}\\
    &\mathcal{L}_2[\vec{\vartheta}, \vec{\eta}] = i  \vec{\vartheta} \cdot \partial_\tau\vec{\eta}  +\tilde{t}' |\delta \vec{\vartheta}|^2+ \tilde{v}' (\partial_\mu \vec{\eta})^2 - \tilde{y}' \sum_i \cos(4 \pi \vec{\alpha}_i \cdot \vec{\eta})\\
    &\mathcal{L}_3[h_0,\vec{\eta}] = - \tilde{y} \sum_i \cos[2\pi \left(h_0 + 2 \vec{\alpha}_i \cdot \vec{\eta}\right)].
\end{split}
\label{eq:eff_lag}
\end{equation}
Here, $\mu$ refers to two spatial coordinate to be summed over and $\delta\vec{\vartheta}\equiv \vec{\vartheta}-\langle\vec{\vartheta}\rangle$, where $\langle\vec{\vartheta}\rangle$ is obtained by minimizing Eq. \ref{eq:H_t}. The cosine terms are present due to the compactness of $h^{r,g,b}$ fields, and we keep the lowest harmonics from combining $\mathcal{V}_{\vec{m}}$'s such that the lattice symmetries are satisfied. 
Here we introduce $\vec{\eta} = (\Re[\eta], \Im[\eta])$ and its corresponding conjugate field $\vec{\vartheta}$. The vectors $\alpha_{1,2,3} = (1,0), (-\frac{1}{2},\frac{\sqrt{3}}{2}),(-\frac{1}{2},-\frac{\sqrt{3}}{2})$ are related by $\mathcal{C}_3$ rotation, where the lattice constant is 1. The cosine terms correspond to $\mathcal{V}_{(1,1,1)} + h.c.$, $\mathcal{V}_{(2,-1,-1)} + h.c. + (\text{cyclic permutation})$, and $\mathcal{V}_{(3,0,0)} + h.c. + (\text{cyclic permutation})$ in $\mathcal{L}_{1,2,3}$ respectively.
The $(\partial_\mu \theta_0)^2$ orignates from ring exchange terms such as $\mathcal{O}_2$.  The $...$ denotes higher-order harmonics and gradient terms. 
What dertermines markedly different fates of $h_0$ and $\vec{\eta}$ is the fact that emergent global quantum number $\tilde{N}$ 
requires the Lagragian to be invariant under $\theta_0 \rightarrow \theta_0 + \text{const}$ and hence forbids the $\theta_0^2$ term, whereas $\delta\vec{\vartheta}^2$ term is allowed by symmetry.

\para After integrating out $\vec{\vartheta}$ in $\mathcal{L}_2$, the action has the sine-Gordon form,
\begin{equation}
 \mathcal{S}_{sg}[\vec{\eta}] =  \int \frac{1}{4 \tilde{t}'^2} (\partial_\tau \vec{\eta})^2 + \tilde{v}' (\partial_\mu \vec{\eta})^2 - \tilde{y}' \sum_i \cos(4 \pi \vec{\alpha}_i \cdot \vec{\eta}).
 \label{eq:eff_eta}
\end{equation}
Let us first ignore $\mathcal{L}_3$. Then we have two decoupled sectors. It is well-known that the sine-Gordon theory in $2+1$D has only a gapped phase, where $\vec{\eta}$ is pinned by the cosine terms. The same mechanism also explains the confinement of compact $U(1)$ gauge theory in $2+1$D and the valence bond solid phases in quantum dimer models on a bipartite lattice in $2+1$D \cite{POLYAKOV1977429}. Here, the two components of the $\vec{\eta}$ map to the two dual photon fields in the compact $U(1) \times U(1)$ gauge theory.
With such $\vec{\eta}$ configurations, $\mathcal{L}_3 \rightarrow -3\tilde{y} \cos{(2 \pi h_0)}$. Therefore, the effect of $\mathcal{L}_3$ is to renormalize $\tilde{y}_0$ to $\tilde{y}_0 + \tilde{y}$ in $\mathcal{L}_1$ to the leading order in $\tilde{y}_0$.

\para Now let us focus on $\mathcal{L}_1$, with $\tilde{y}_0 \rightarrow \tilde{y}_0 + \tilde{y}$. We note that $\mathcal{L}_1$ resembles the effective action for quantum interface with continuous symmetry \cite{fradkin1983roughening}. The difference is that the height field in $\mathcal{L}_1$ is compact, $h_0 \sim h_0 + 3$, while in the quantum interface, the height is not compact. 

 As pointed out by Ref.\cite{fradkin1983roughening}, the cosine term $\cos{2 \pi h_0}$ is relevant compared to the $(\partial_\mu h_0)^2$ term. Hence, the long-wavelength properties is governed by fluctuation of the height fields around the minima of the cosine term,
\begin{equation}
    \mathcal{L}_{1,n_0}[\theta_0, h_0] \approx i \theta_0 \partial_\tau h_0 + \tilde{t} (\partial_\mu \theta_0)^2 + \tilde{v}_0(\partial_\mu h_0)^2 + 2\pi^2 \tilde{y}_0 (\delta h_0)^2,
    \label{eq:hydro}
\end{equation}
where $\delta h_0  = h_0 - n_0$ and $n_0$ is an integer. The path integral takes the sum over all the minima, $\mathcal{Z}_1 \sim  \int D h_0 D\theta_0 \sum_{n_0} e^{- \int \mathcal{L}_{1,n_0}[\theta_0, h_0] }$. 
Ref.~\cite{fradkin1983roughening} essentially integrated out $\theta_0$ and focused on the physics of $\delta h_0$. However, the action involving $\mathcal{L}_1$, for each $n_0$, can be viewed as the hydrodynamic action of a bosonic superfluid in the long-wavelength limit by mapping $\theta_0$ to the phase angle of the condensate \cite{popov1972hydrodynamic, Popov_1988}. Hence, we identified the SN phase. The ordering of $\vec{\eta}$ that is guaranteed in the ground state from analyzing $\mathcal{S}_{sg}[\vec{\eta}]$ (Eq.\ref{eq:eff_eta}), irrespective of spontaneous symmetry breaking or lack thereof in the theory of $\theta_0$, accounts for the ``nematic'' part of the phase. The ordering of $\theta_0$ accounts for the `` super'' part in analogy to superfluid, although this is a phase conjugate to the total sublattice polarization rather than total particle number, hence SN is not a superfluid.
The $\theta_0$-disordered phase will be analogous to the Mott insulating state in the Bose-Hubbard model \cite{Fisher1989Phys.Rev.B}, which is a $ C_3$-symmetry-breaking, columnar-ordered phase in our case. Hence, we established the two possible ground state phases in the phase diagram sketched in Fig.\ref{fig:model} e).

Specifically, the two distinct ground states are:
\begin{itemize}
\item 
$\langle e^{i \frac{2\pi}{3} h_0} \rangle \neq 0,~ \langle e^{i \frac{4\pi}{3} \vec{\alpha}_i \cdot \vec{\eta}} \rangle \neq 0$: the gapped \textbf{columnar ordered phase}.
\item 
$\langle e^{i \theta_0 }\rangle \neq 0,~ \langle e^{i \frac{4\pi}{3} \vec{\alpha}_i \cdot \vec{\eta}} \rangle \neq 0$: the gapless \textbf{supernematic phase}.
\end{itemize}
We discuss the properties of each phase in the following sections.

A complete theory of the quantum phase transition between the two phases requires addressing several subtleties, which are beyond the scope of the present paper. The central issue is that the vertex operators $e^{i\alpha h_0}$ and $e^{i p \theta_0}$ do not commute. As a result, the symmetry of $\mathcal{L}_1$ takes the nontrivial form $U(1)\rtimes Z_3$, with group elements $e^{i\alpha h_0}$ ($\alpha\in[0,2\pi)$) and $e^{i p \theta_0}$ ($p\in{0,1,2}$), respectively. The impossibility of simultaneously ordering these two vertex operators introduces an additional subtlety that was previously unnoticed. Formally integrating out the $h_0$ field in Eq.\ref{eq:hydro} as one would in the hydrodynamic theory of superfluid yields the phase-only action:
\begin{equation}
    \mathcal{S}_1[\theta_0] = \int \frac{1}{\tilde{\mathbf{y}}}(\partial_\tau \theta_0 )^2 + \tilde{t}(\partial_\mu \theta_0 )^2+ ...,
    \label{eq:eff_theta}
\end{equation}
where $\tilde{\mathbf{y}}$ is a monotonic function of $\tilde{y}_0 + \tilde{y}$. The Euclidean action $\mathcal{S}_1[\theta_0]$ is analogous to the 3D classical XY model with $\theta_0$ as the spin azimuthal angle. 
If the vortices of $\theta_0$ can be taken into account the same way as in the XY model, there would be a phase transtion between a superfluid phase and a gapped Mott insulating phases by tuning the effective stiffness $K_{\text{eff}} = \sqrt{ \tilde{t}/\tilde{\mathbf{y}}}$ across a critical value $K_{\text{eff}}^c$\cite{WilsonFisher}. We leave the full theory of the phase transition upon introducing vortices of $\theta_0$ to a future work.

\section{Columnar ordered phase}
\label{sec:CO}
\para In the classical limit of $\tilde{t} = 0$, we reproduce the trimer liquid\cite{zhang2024bionicfractionalizationtrimermodel}, a classical ensemble of extensively-many degenerate configurations (\ref{fig:model} e)). In general, small quantum fluctuations around extensive degeneracy can result in quantum order-by-disorder, where a charge ordering state is favored due to the virtual exchange processes. To determine the symmetry-breaking pattern of such a charge-ordered state and its thermal behavior, we use the effective field theory and the symmetry representations of the fields.

\subsection{Zero temperature}
\para The order parameters of the columnar ordering phase are the lowest order vertex operators, with the following expectation values:
\begin{equation}
    \begin{split}
        \langle \mathcal{V}_{(1,0,0)} \rangle &= e^{i \frac{2\pi}{3}(\bar{h}_0 + e_1)}\\
        \langle \mathcal{V}_{(0,1,0)} \rangle &= e^{i \frac{2\pi}{3}(\bar{h}_0 + e_2)}\\
        \langle \mathcal{V}_{(0,0,1)} \rangle &= e^{i \frac{2\pi}{3}(\bar{h}_0 - e_1 - e_2)},
    \end{split}
    \label{eq:V_avg}
\end{equation}
where $\langle h_0 \rangle = \bar{h}_0 $  and $\vec{\eta} = \left(\frac{e_1}{2}, \frac{\sqrt{3}}{3}(\frac{e_1}{2}+ e_2) \right)$, where $\bar{h}_0$, $e_{1,2}$ are integers. There are three distinct values for $\langle \mathcal{V}_{(1,0,0)} \rangle$ and we find $\langle \mathcal{V}_{(0,1,0)} \rangle$ respectively, and $\langle \mathcal{V}_{(0,0,1)} \rangle = \langle \mathcal{V}_{(1,0,0)} \rangle^* \langle \mathcal{V}_{(0,1,0)} \rangle^*$ from Eq.\ref{eq:V_avg}. Therefore, there are nine distinct charge ordering phases in total, which correspond to the columnar ordering in Fig.\ref{fig:model} c) and its symmetry-related counterparts.
Physically, the $\langle \mathcal{V}_{(1,0,0)} \rangle$, $\langle \mathcal{V}_{(0,1,0)} \rangle$, and $\langle \mathcal{V}_{(0,0,1)} \rangle$ correspond to the order parameter of the columnar ordering of the red, green, and blue double-dimers, respectively. The charge ordering gives rise to Bragg peaks at momentum $K_{CO} = \frac{4\pi}{3\sqrt{3}}(1,0)$ and all the $C_6$ related momenta. The CO state is a ``Mott-insulator'' of $\tilde{N}$ that breaks lattice translational symmetry due to $\langle h_0\rangle\neq0$. It also breaks the threefold rotational symmetry of the lattice, $C_3$, due to  $\langle \vec{\eta}\rangle\neq0$.

\subsection{Finite temperature}
\label{sec:finite_CO}
\para To effective field theory Eq.\ref{eq:eff_lag} can be extended to finite temperature. Deep in the CO phase, the effective action is a 2D sine-Gordon theory with three components,
\begin{equation}
    \begin{split}
        S_{CO}[h_0,\eta] = \frac12 K_0 \int (\partial_\mu h_0)^2 + \frac12 K' \int (\partial_\mu \vec{\eta})^2  -\sum_{\vec{m}} y_{\vec{m}} \mathcal{V}_{\vec{m}},
    \end{split}
    \label{eq:S_CO}
\end{equation}
where $K_0 = 2\beta \tilde{v}_{0,\text{eff}}$, $K' = 2\beta \tilde{v}_{\text{eff}}'$. 
We note that, from Eq.\ref{eq:eff_lag}, for the ordering of $h_0$, the corresponding cosine term is given by $\mathcal{V}_{(1,1,1)}$, whose scaling dimension is $\Delta_{(1,1,1)} = \frac{\pi}{K_0}$. For the ordering of $\vec{\eta}$, the corresponding term is $\mathcal{V}_{(2,-1,-1)}$, whose scaling dimension is $\Delta_{(2,-1,-1)} = \frac{4\pi}{K'}$. Hence, at low temperatures, all cosine terms are relevant around the Gaussian fixed point, leading to the CO. 

\para To understand thermal melting of the CO we need to consider 
thermally excited violations of the cluster-charging interaction. A change in the occupation number at a single site violates three cluster-charging constraints associated with the three plaquettes sharing the site. Such excitation carries two effects as follows. (1) Violation of cluster-charging of a single plaquette generates ``bionic" monomers, i.e., monomers of two colors of the height fields, as discussed for the classical trimer model at filling $1/3$ \cite{zhang2024bionicfractionalizationtrimermodel}. The monomer charges of $(h^r, h^g, h^b)$-fields are $(0,3,-3)$, $(-3,0,3)$ and $(3,-3,0)$ for the bionic monomers residing in the red, green and blue plaquettes in Fig.\ref{fig:double-dimer} b). Hence removing or adding one particle creates a bound state of three bionic monomers, where the total monomer charges sum to zero.
(2) From the relation Eq.\ref{eq:p_duality}, changing boson number of one single site corresponds to shifting $h_0$ by $1$ after coarse-graining. The shift at site $x_0$ can be viewed as a ``charge" of the $\vec{\nabla} h_0$ field and the configuration of $h_0$ satisfies $\nabla^2 h_0(x) = \delta^2(x-x_0)$. Hence, the scaling dimension of this shift operator is $\Delta_s = \frac{K_0}{4\pi}$, following the same argument as in the vortex operator in the XY model \cite{kosterlitz1973ordering}. The critical $K_0$ for the shift is $K_{0,c} = 8\pi$ while the cosine term scales as $\Delta_{(1,1,1)}(K_{0,c}) = \frac{1}{8} <2$ and is relevant. Hence, the thermal phase transition is dominated by the cosine term $\cos{2\pi h_0}$ \cite{PhysRevB.16.1217}. Since $h_0 \sim h_0 +3$, the ordering of $h_0$ is three-fold degenerate, and we expect the thermal transition of $h_0$ to be three-state Potts-like.

\begin{figure}
    \centering
    \includegraphics[width=0.95\linewidth]{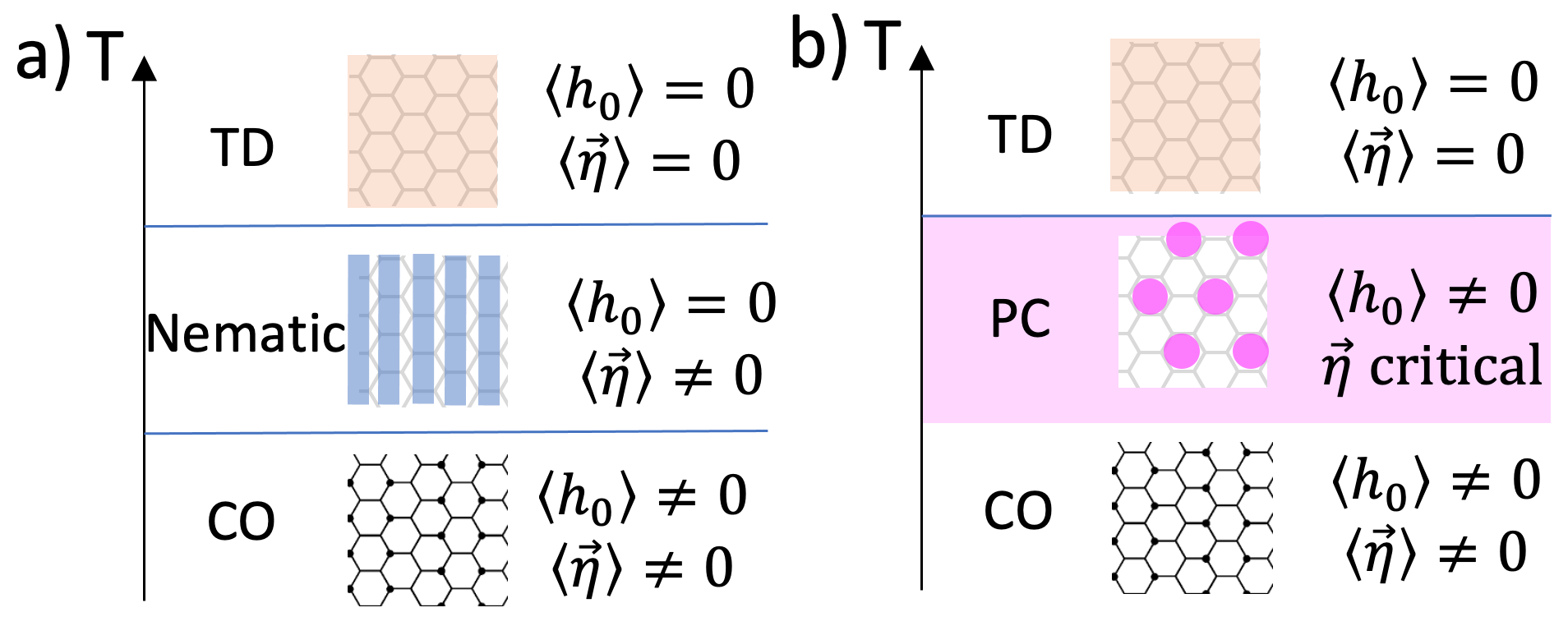}
    \caption{Two scenarios of the thermal phase transition of the Columnar order, with the cartoon picture denoting the ordering patterns. The nematic order breaks $\mathcal{C}_3$ while preserving translation and inversion. PC: palquette Coulomb, $\sqrt{3}\times\sqrt{3}$ quasi-long range ordering while preserving $C_6$ rotation. TD: thermally disordered. The blue lines denote a three-state Potts transition. The magenta region denotes quasi-long range order.}
    \label{fig:thermal_CO}
\end{figure}

\para In general, the ordering of $h_0$ and $\vec{\eta}$ do not coincide. The bionic monomers can be viewed as vortices of the $\vec{\eta}$-field. Since the physical boson couples the shift in $h_0$ to the three-body bound state of the bionic monomers, the bionic monomers are always confined in the ordering phase of $h_0$. This is because, to dissociate the bionic monomers, one needs to shift the value of $h_0$ along a line, yielding a linear energy cost. Therefore, when $\langle h_0 \rangle \neq 0$, the monomer fugacity for $\vec{\eta}$ is always zero at large distance. Hence, depending on the relative strength of $K_0$ and $K'$, there are two scenarios for the thermal melting of the CO, illustrated in Fig. \ref{fig:thermal_CO}:

\para (a) Upon increasing temperature, $h_0$ melts first, via a three-state Potts transition. The intermediate phase has $\vec{\eta}$ ordering. Since $h_0$ is disordered, the order parameters are the lowest order vertex operators involving only $\vec{\eta}$-field: $\langle \mathcal{V}_{(1,-1,0)} \rangle$, $\langle \mathcal{V}_{(0,1,-1)} \rangle$ and $\langle \mathcal{V}_{(-1,0,1)} \rangle$. They all take the same value of $1,e^{i \frac{2\pi}{3}},\text{or} e^{-i \frac{2\pi}{3}}$ from Eq.\ref{eq:V_avg}. From Eq.\ref{eq:V_sym}, the ordering pattern breaks $\mathcal{C}_3$ rotation while preserving translation and inversion. Hence, this intermediate phase is three-fold degenerate, which we call ``nematic".
In the disordered phase of $h_0$, the bionic monomers can acquire non-zero fugacity. The scaling dimension of these monomer operators can be obtained from the same method as in the classical trimer liquid \cite{zhang2024bionicfractionalizationtrimermodel} and $\Delta_{m_\eta} = \frac{3 K'}{4\pi}$. Hence when $K' = K_{c}^{'} = \frac{8\pi}{3}$, the monomers start to become relevant. However, since $\Delta_{(2,-1,-1)}(K_c^{'}) = \frac{3}{2} < 2$, the thermal phase transition is dominant by the cosine terms, where the value of the critical $K'$ is nonuniversal, and expected to be three-state Potts like \cite{PhysRevB.16.1217}.

\para (b) Upon increasing temperature, $\vec{\eta}$ melts first while $h_0$ is ordered. Since the bionic monomers have zero fugacity in the ordering phase of $h_0$, the melting of $\vec{\eta}$ is driven by the cosine terms becoming irrelevant, which happens at $K' \equiv K_{\text{lower}}' = 2\pi$ (lower boundary of the magenta region in Fig.\ref{fig:thermal_CO} b)). Above this temperature, the effective theory of $\vec{\eta}$ is Gaussian and critical. This phase has $C_6$ rotation symmetry and the three-sublattice translation breaking induced by $\langle h_0 \rangle$ is quasi-long range, which we call ``plaquette Coulomb phase". Further increases in temperature cause the three-state Potts transition of melting $h_0$. Since the bionic monomer at most has scaling dimension $\frac{3 K_{\text{lower}}'}{4 \pi} = \frac{3}{2} < 2$, the Coulomb phase terminates at the thermal melting transition of $h_0$, where bionic monomers proliferate, and the stiffness $K'_{\text{upper}}$ right below the transition (upper boundary of the magenta region in Fig.\ref{fig:thermal_CO} b)) can related to the lower boundary of the PC phase as $K'_{\text{upper}} = 2\pi T_1 /T_2$, where $T_{1,2}$ are the lower and upper temeprature of the PC phase, from the Gaussian field theory.  

\section{Supernematic}
\label{sec:SN}
\begin{figure*}
    \centering
    \includegraphics[width=0.95\linewidth]{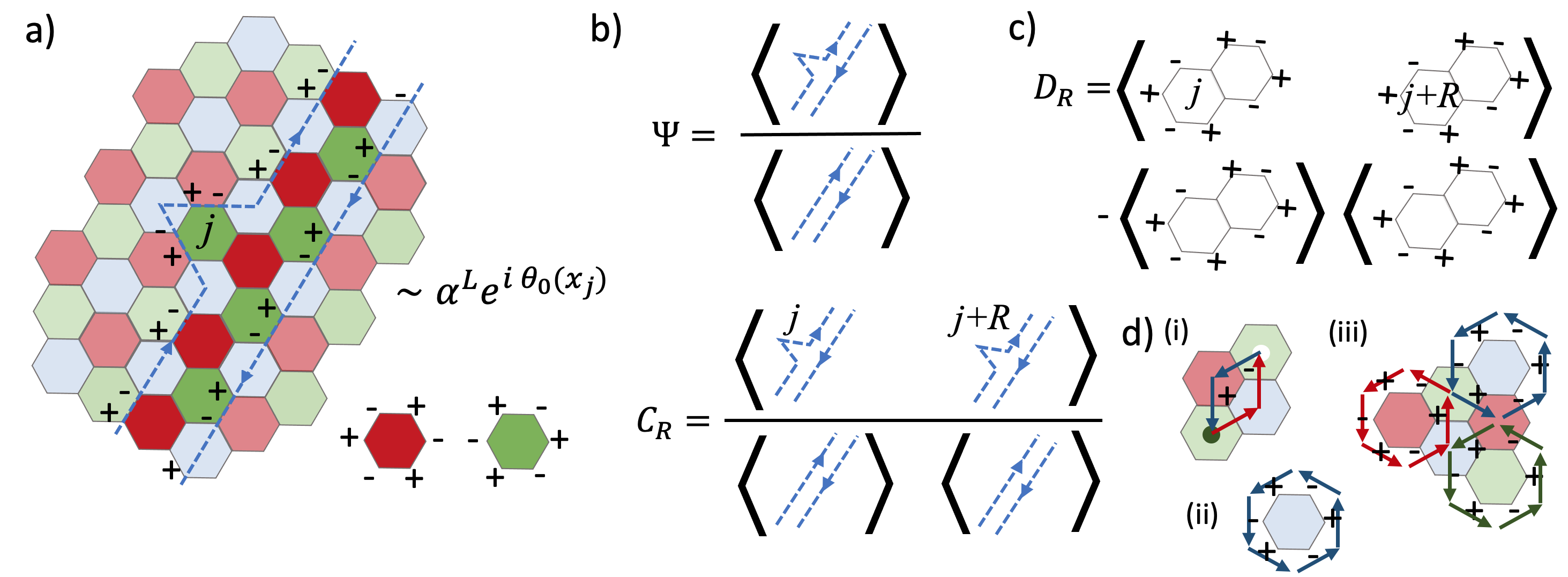}
    \caption{a) Example of a charged operator under $N_A-N_B$. The $\pm$ denote creation and annihilation operators of the particles, and the arrows denote two different assignments of the consecutive $\pm$'s. Note that we have colored the honeycomb plaquette with three colors and demand that the double loops consist of lines traversing plaquettes of the same color (blue in the figure). 
    %This requirement is to ensure that the tilt of the height fields do not change so that the operator is local in the long wavelength limit (see Appendix for more detailed discussion). 
    Bottom: sign assignment for the red and green plaquettes.  b) Order parameter and off-diagonal long-range order defined on the lattice. c) Connected correlator of $\mathcal{O}_2$'s with seperation $R$. d) (i) Bionic monomer hopping from the green plaquette to another green one. (ii) Single plaquette flipping process assisted by the presence of one bionic monomer. (iii) Lowest order symmetry breaking process involving only $\theta_0$-fields at finite temperature. The solid and hollow circles denote increasing and decreasing monomer numbers, respectively. The colored arrows denote the gauge field capturing the Gauss Law of the cluster-charging constraints.
    }
    \label{fig:obs}
\end{figure*}
\subsection{Zero temperature}
\para The low-energy description above establishes SN as a ``superfluid'' of sublattice particle number $\tilde{N} = N_A - N_B$, the new quantum number for the cluster-charging model based on the Conway-Lagarias invariant. 
From the perspective of the quantum number $\tilde{N}$, properties of SN should follow from well-known properties of a superfluid. However, total particle number $N = N_A + N_B$ is gapped, rendering SN incompressible. Hence, SN is an unusual new state of matter that is macroscopically coherent, yet incompressible. Moreover, since the tiling invariant $\tilde{N}$ cannot be changed locally, the order parameter defined on the lattice has to be nonlocal.
The SN has a unique ``intertwined order'' that breaks the global $U(1)$ symmetry associated with $\tilde{N}$ conservation due to $\langle \theta_0 \rangle \neq 0$ and three-fold rotation $C_3$ due to pinning of $\vec{\eta}$ as we show below.

\para %In the SN, $\langle \theta_0 \rangle \neq 0$.
Since $h_0$ is fluctuating in the SN, the lowest order vertex operators $\mathcal{V}_{(1,0,0)}$'s do not acquire expectation values, which means that each single color in the double-dimer model is not ordered. However, the joined configuration of different colors might have ordering due to the pinning of the $\eta$-field. To address this question, we consider vertex operators involving only $\vec{\eta}$-fields. The lowest order operators are $\mathcal{V}_{(1,-1,0)}$, $\mathcal{V}_{(0,1,-1)}$, and $\mathcal{V}_{(-1,0,1)}$. All the other $\vec{\eta}$-only vertex operators can be generated by considering the complex conjugation and multiplication of the above ones. The vertex operator $\mathcal{V}_{(1,-1,0)}$ corresponds to the multiplication of the dimer densities of the neighboring red and green dimers, similarly for the $\mathcal{V}_{(0,1,-1)}$, and $\mathcal{V}_{(-1,0,1)}$. 
The expectation values of these vertex operators are:
\begin{equation}
    \begin{split}
        \langle \mathcal{V}_{(1,-1,0)} \rangle &= e^{i \frac{2\pi}{3}(e_1 - e_2)}\\
        \langle \mathcal{V}_{(0,1,-1)} \rangle &= e^{i \frac{2\pi}{3}(e_1 + 2 e_2)}\\
        \langle \mathcal{V}_{(-1,0,1)} \rangle &= e^{i \frac{2\pi}{3}(- 2e_1 - e_2)},
    \end{split}
    \label{eq:VSN_avg}
\end{equation}
where $\vec{\eta} = \left(\frac{e_1}{2}, \frac{\sqrt{3}}{3}(\frac{e_1}{2}+ e_2) \right)$, the same as in the CO case. We note that $\langle \mathcal{V}_{(1,-1,0)} \rangle = \langle \mathcal{V}_{(0,1,-1)} \rangle = \langle \mathcal{V}_{(-1,0,1)} \rangle$, and can take value of $1, e^{i \frac{2\pi}{3}},\text{or } e^{-i \frac{2\pi}{3}}$. This ordering breaks $\mathcal{C}_3$ but preserves inversion and translation. Hence, we can label the SN phase at zero temperature by two order parameters, $\langle e^{i \theta_0}\rangle$ and $\langle \mathcal{V}_{(1,-1,0)} \rangle$. Since $\theta_0 \rightarrow - \theta_0$ under inversion, the SNs with $\langle e^{i \theta_0}\rangle = \pm 1$ are inversion symmetric, and are otherwise inversion breaking.

\para
As a ``superfluid'' of $\tilde{N}$, the SN has a gapless Goldstone mode with the corresponding dynamical structure that is linear in $q$. Specifically, we can 
express the particle density in terms of the coarse grained fields, 
\begin{equation}
    n_A(x) - \frac{1}{3} =  - n_B(x) + \frac{1}{3} \sim -\frac{1}{2\pi}\sin{2\pi h_0} + ...,
\end{equation}
where we ignore the dependence on the gapped $\eta$-fields and a spatially oscillating term (see SM \ref{sec:app:h} for the derivation).
Following the effective field theory of $\theta_0$ (Eq.\ref{eq:eff_theta}) and the standard derivation following the hydrodynamic theory of bosonic superfluid \cite{Popov_1988} (see SM \ref{sec:sf}), the dynamical structure factor $\tilde{S}(q,\omega) \sim |q| \delta(\omega - v|q|)$, where $\tilde{S}(q,\omega) = \int dt e^{i \omega t}\langle \tilde{n}(q,t) \tilde{n}(-q,0)\rangle$, and $\tilde{n}(q)$ is the Fourier transform of $n_A(x_A)- n_B(x_B)$. Therefore, the SN is compressible with respect to the sublattice particle number $\tilde{N}$, following the compressibility sum-rule \cite{10.1093/acprof:oso/9780198758884.003.0007} $\tilde{\kappa} =  2 \lim_{q\rightarrow 0} \int_0^\infty \frac{\tilde{S}(q,\omega)}{\omega} d\omega$. Hence, from the $\tilde{N}$ perspective, SN is compressible although it is an incompressible state for the total particle number $N=N_A+N_B$. 

\para Despite the seeming resemblance of the effective field theory of the SN Eq.~\eqref{eq:eff_theta}, to that of a typical superfluid, the SN carries the fingerprint of its unusual microscopic origin tied to the global CL tiling invariant through a \emph{non-local} order parameter. 
Unlike an ordinary superfluid, the SN has no local order parameter because $\tilde N$ is exactly conserved, i.e, a local operator charged under $\tilde N$ is forbidden. 
Instead, we find a non-local string operator that carries $\tilde N$ charge and reduces to the local field $e^{i\theta_0}$ in the long-wavelength limit. To construct this charged string operator, first consider a single hexagonal plaquette that hosts a height field. An operator consisting of  
alternating creation and annihilation operators around this plaquette (the green plaquette in Fig. \ref{fig:obs}a)) changes the corresponding height by $3$ (and hence changes $\tilde{N}$), but annihilates all states in the constrained Hilbert space, because it demands three particles on one plaquette. However, we can stay within the constrained Hilbert space by considering an extended operator that stitches plaquettes supporting two of the three height-field colors (say red and green) along a non-contractible loop (Fig. \ref{fig:obs}a)). Adding one extra hexagon -- the ``bump” -- to this extended operator ensures a net change in $\tilde N$. The resulting operator is supported on two oppositely oriented lines connecting blue plaquettes (blue dashed lines in Fig. \ref{fig:obs}a)), where the orientation tracks the ordering of creation and annihilation operators. \footnote{The presence of the boundary invariant requires particular care when taking the thermodynamic limit. 
Under periodic boundary conditions, the total number of unit cells along one direction must be a multiple of three to realize the exact commensurate fillings and to define the charged loop operator. 
For open boundaries, the edges must be softened to enable access to different $\tilde{N}$ sectors, and the charged operator would be touching the boundary. }

\para Although this $\tilde{N}$-charged operator is loop-like, the $\tilde{N}$ symmetry differs fundamentally from the 1-form symmetry in gauge theories: the $\tilde{N}$ charge of the double-loop depends on its shape, in contrast to the Wilson loop in gauge theory, whose charge depends only on topological data. In the continuum, a double-loop with a bump around plaquette $x_j$ matches $\alpha^{L} e^{i\theta_0(x_j)}$ to leading order, where $L$ is the string length and $\alpha$ is a non-universal perimeter-law factor arising from integrating out the gapped $\eta$ sector (and from the short-distance contribution to $e^{i\oint \nabla\theta\cdot dl}$). 
We can define a normalized expectation value of the ``bump'' operator, $\Psi$, as the ratio between the expectation values of the double loop with and without the bump, as shown in Fig. \ref{fig:obs} b). We also define an analogously normalized ``bump'' correlation function for the ``bump''s separated by distance $R$, $C_R$. Now one can detect the SN phase
by observing a nonzero $\Psi$ under infinitesimal symmetry breaking and by a finite $C_R$ at large separation $R$. This corresponds to a \emph{non-local}, off-diagonal, long-range order (ODLRO). In finite-size simulations, connected correlators of local operators, such as density or $\mathcal{O}_2$ (Fig. \ref{fig:obs}c), decay as a power law in the SN phase but exponentially in the charge-ordering phase, providing a practical diagnostic.

\begin{figure}
    \centering
    \includegraphics[width=0.95\linewidth]{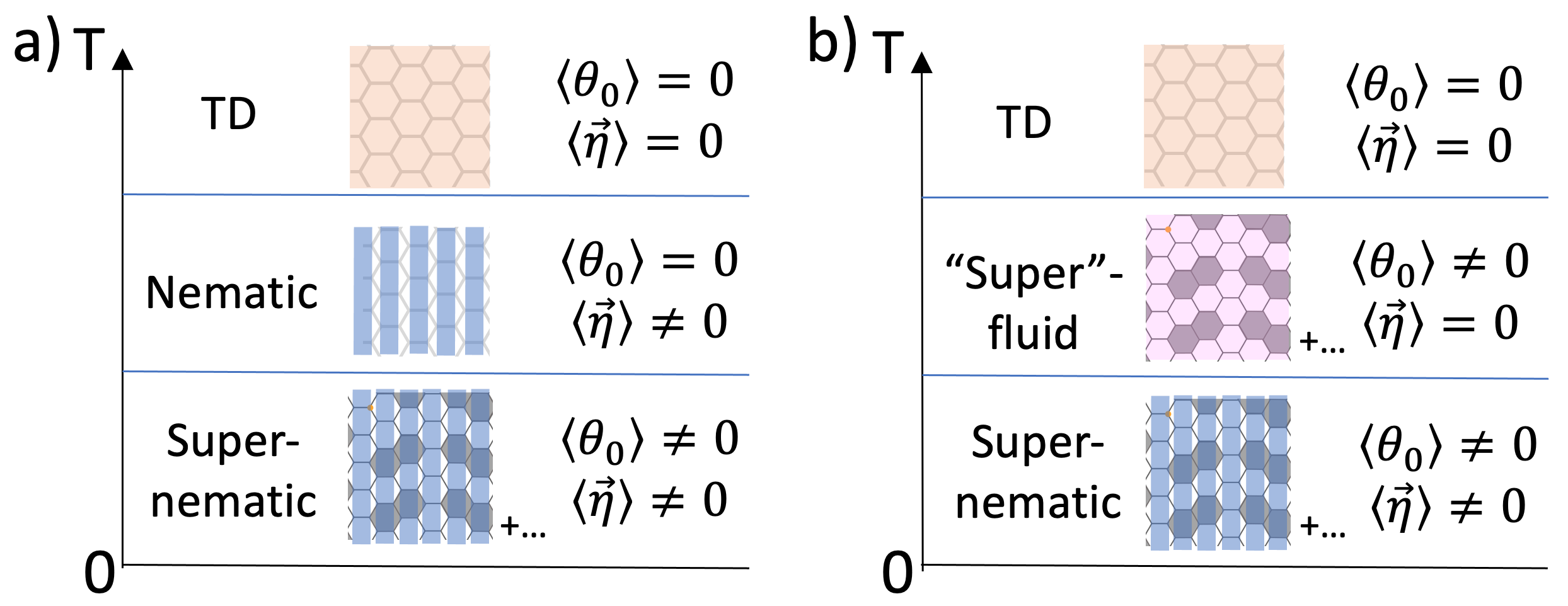}
    \caption{Two scenarios of the thermal phase transition of the super-nematic phase, with the cartoon picture denoting the ordering patterns. TD: thermally disordered. The blue lines denote three-state Potts transition.}
    \label{fig:thermal_SN}
\end{figure}

\begin{figure*}
    \centering
    \includegraphics[width=0.95\linewidth]{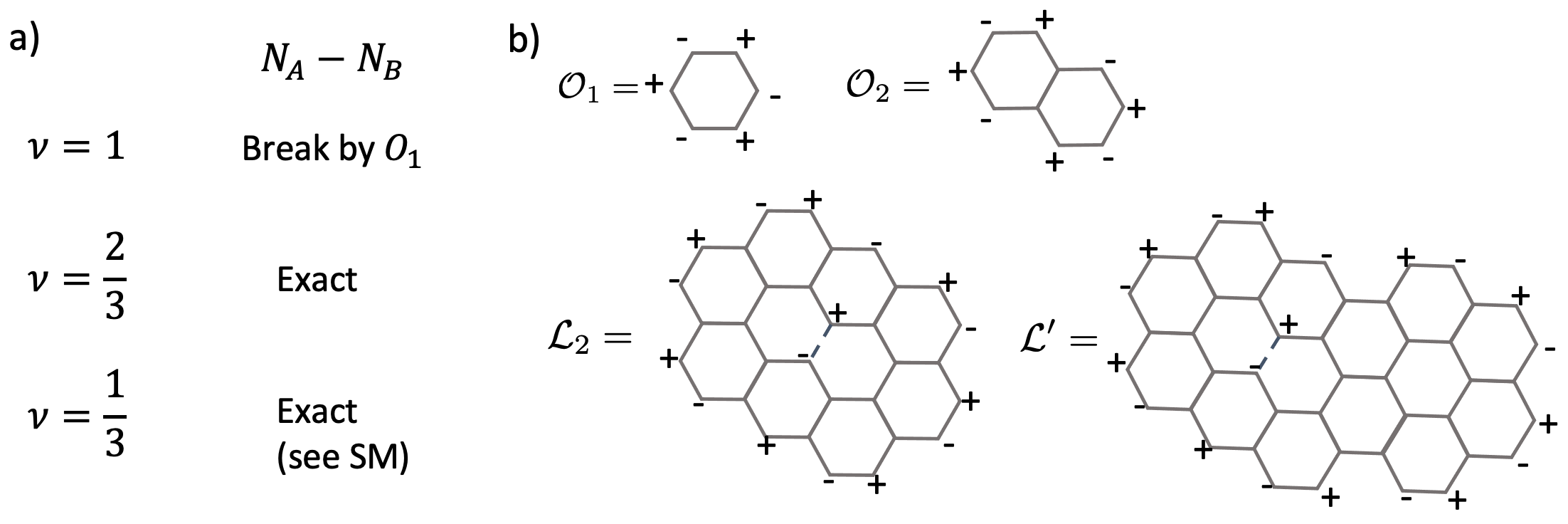}
    \caption{a) The presence or absence of the $\tilde{N}= N_A - N_B$ conservation. Here exact means present upto all local operators. b) Some non-zero ring exchange operators. $\mathcal{O}_1$, $\mathcal{O}_2$ and $\mathcal{L}_2$ are the lowest order operators for filling $1, 2/3$ and $1/3$ respectively. $\mathcal{L}_2$ also conserves the dipole moment of $\tilde{N}$, which is broken by higher order operators such as $\mathcal{L}'$.}
    \label{fig:other-filling}
\end{figure*}

\subsection{Finite temperature}
\label{sec:finite_SN}

\para At finite temperature, the $\tilde{N}$ symmetry is no longer exact, since thermally activated bosonic hopping between the $A$ and $B$ sublattices allows local processes that change $\tilde{N}$. The central question is therefore whether the finite-temperature phase of the SN can still be viewed as a \emph{superfluid} of $\tilde{N}$, despite the presence of these thermally induced violations. To address this, we derive the leading $\tilde{N}$-breaking terms generated by thermal fluctuations.

\para In the imaginary-time path-integral language, the thermal partition function can be written as,
\begin{equation}
    Z = \Tr e^{-\beta H}
    \sim \int \mathcal{D}[\theta_0, \vec{\eta},m_\alpha]\,
    e^{-S_E[\theta_0, \vec{\eta},m_\alpha]},
\end{equation}
where $m_\alpha$ denotes the monomer fields, to be specified later. We decompose the action as $S_E[\theta_0, \vec{\eta},m_\alpha] = S_0 [\theta_0, \vec{\eta}] + S_{\text{viol}}$, where $S_0$ is the thermal action for the constrained Hilbert space and $S_{\text{viol}}$ denotes the contribution from the thermal violation of the cluster-charging constraints.

\para $S_0$ is governed by the effective action obtained by taking the classical limit of our quantum effective theory (ignoring the $\tau$-dependent terms in Eqs.~\ref{eq:eff_eta} and \ref{eq:eff_theta}),
\begin{equation}
    S_0[\theta_0, \vec{\eta}] = \int d^2x \left(
    \frac{1}{2}\rho_0 (\partial_\mu \theta_0)^2 
    + \frac{1}{2}K' (\partial_\mu \vec{\eta})^2 
    - \sum_{\vec{m}} y_{\vec{m}} \mathcal{V}_{\vec{m}}
    \right),
\end{equation}
where $\rho_0 = 2\beta \tilde{t}$ is the phase stiffness, and $\mathcal{V}_{\vec{m}}$ denotes the vertex operators. The $\vec{\eta}$ sector is described by a two-component sine-Gordon theory identical to that in the CO phase (Eq.~\ref{eq:S_CO}).

\para $S_{\text{viol}}$ consists of two contributions: the monomer hopping term and the cluster-charging energy,
\begin{equation}
    S_{\text{viol}} = \int_0^\beta d\tau \sum_{\alpha, i,j} -t m_{\alpha,i}^\dagger\, 
    e^{i (A_{ij}^\delta - A_{ij}^\gamma)} m_{\alpha,j} + \sum_{\alpha, i} U m_{i,\alpha}^\dag m_{i,\alpha}.
\end{equation}
Here we only consider monomer hopping terms mapped to the boson hopping between nearest neighbors, $-t b_i^\dagger b_j$. The higher-order hopping terms can be written in a similar way. $m_{\alpha,i}$ denotes the annihilation operator for a ``bionic'' monomer residing at plaquette center $i$, and $\alpha,\delta,\gamma = r,g,b$ label the color indices (see Fig.~\ref{fig:obs} d)(i)) with the following rule. A monomer $m_{g,i}$ located on a green plaquette violates the cluster constraint on that plaquette and, consequently, the dimer constraints on both the red and blue bonds in the corresponding dimer representation, similarly for $m_{r,i}$ and $m_{b,i}$. 
Interpreting the dimer constraints as Gauss laws, $m_{g,i}$ carries opposite gauge charges under the emergent fields $A^r$ and $A^b$ (defined on different links as shown in Fig.~\ref{fig:obs} d)(i)), realizing the ``bionic'' character introduced in Ref.~\cite{zhang2024bionicfractionalizationtrimermodel}.  The magnetic flux defined by summing the gauge fields $A_{ij}^\alpha$ around a hexagonal plaquette centered at site $l$ is related to the conjugate height field as 
$\sum_{\langle ij \rangle \in \hexagon_l} A_{ij}^\alpha = 3\,\theta_l^\alpha$. 

\para The effect of thermally generated monomers can be captured by integrating over all possible monomer configurations. 
This procedure yields gauge-invariant combinations of the link fields $A_{ij}$'s, corresponding to ring-exchange processes mediated by virtual monomer excitations. 
Owing to the bionic nature of the monomers, the lowest-order gauge-invariant process necessarily involves three successive monomer hoppings around a hexagon, resulting in a three-particle flip on a single plaquette (Fig.~\ref{fig:obs}d)(ii)), which generates the term
\begin{equation}
   -w_1 e^{-\beta U} 
   \sum_{\alpha = \{r,g,b\}} \sum_l 
   \cos(3\theta_l^\alpha),
   \label{eq:assis_ring1}
\end{equation}
where $w_1 \sim O(t^3/U^2)$, and the Boltzmann factor $e^{-\beta U}$ accounts for the probability that plaquette $l$ hosts three particles, enabling the flip to occur.

\para Eq.\ref{eq:assis_ring1} explicitly breaks $\tilde{N}$ conservation since it is not invariant under constant shift of $\theta_0$. The question now is whether it destroys the SN phase at zero temperature. $\theta^{\alpha = r,g,b}$'s are linear combinations of $\theta_0$ and $\vec{\vartheta}$ fields. In the ordered phase of $\vec{\eta}$, the fields $\vec{\vartheta}$ are pinned. They acquire expectation values that minimize Eq.~\ref{eq:H_t}. 
The minima satisfy $(3\theta^r - 3\theta^g,\, 3\theta^g - 3\theta^b,\, 3\theta^b - 3\theta^r) 
= 2\pi (n_1, n_2, -n_1 - n_2)$ with integers $n_{1,2}$, corresponding to $(\vartheta_x, \vartheta_y) 
= 2\pi (-n_2/\sqrt{3},\, (2n_1 + n_2)/3)$. 
Pinning of the $\vartheta$ fields induces cosine terms in $\theta_0$ via the assisted single-plaquette flip (Eq.~\ref{eq:assis_ring1}).
This term is relevant in the superfluid phase of $\tilde{N}$, and its minima satisfy 
$\cos(\theta_0 + 2\pi (2n_1 + n_2)/3) = 1$ for $t>0$. 
However, summing over all degenerate vacua of the $\vec{\eta}$ ordering—i.e., all integer pairs $(n_1,n_2)$—renders $\theta_0$ effectively a random angle that can be any integer multiples of $2\pi/3$. 
On average, Eq.~\ref{eq:assis_ring1} therefore does not fix $\theta_0$.

\para To capture the leading term that directly couples to $\theta_0$, we consider ring-exchange processes independent of $\vec{\vartheta}$. 
The dominant contribution is a three-plaquette process (Fig.~\ref{fig:obs}d)(iii)) that produces a threefold anisotropy,
\begin{equation}
   - w_3 e^{-3\beta U} \int d^2x\, \cos(3\theta_0),
\end{equation}
with $w_3 \sim O(t^9/U^8)$. 
This term explicitly breaks the $\tilde{N}$ symmetry and pins $e^{i\theta_0}$ to one of the three minima: $1$, $e^{i2\pi/3}$, or $e^{-i2\pi/3}$.

\para Therefore, the ordering of the phase angle $\theta_0$ is discrete and long-ranged, characterized by a $Z_3$ order. 
The connected two-point correlator exhibits a finite correlation length, distinguishing the thermal phase of the SN from a genuine superfluid phase of $\tilde{N}$ in the thermodynamic limit. Nevertheless, the effect of the threefold anisotropy can remain weak over a large length scale, owing to the small Boltzmann factor $e^{-3\beta U}$ and the high-order hopping amplitude $w_3$. 
The length scale beyond which the anisotropy becomes relevant defines a pinning length $l_p$. 
Within the range of $l_p$, the thermal SN phase effectively behaves as a superfluid of $\tilde{N}$. 
To leading order in $w_3$, the pinning length is estimated from the renormalization group analysis\cite{kosterlitz1973ordering} as,
\begin{equation}
    l_p \sim \left(w_3^{-1} e^{3\beta U}\right)^{\frac{1}{2-\Delta_3}} > e^{3\beta U/2},
\end{equation}
where $\Delta_3 = 9 / (4\pi \rho_0)$ is the scaling dimension of the $\cos(3\theta_0)$ operator. 
Despite the explicit anisotropy, the exponential factor ensures that $l_p$ can remain exceedingly large. 
For instance, at $T/U \sim 0.1$, the pinning length exceeds $10^6$ lattice spacings, indicating that the zero-temperature discussion based on exact $\tilde{N}$ conservation remains effectively valid even at finite temperature.

\para Now let us discuss the thermal melting of the SN. Because both $\theta_0$ and $\vec{\eta}$ order in SN, it can melt in two ways (Fig.\ref{fig:thermal_SN}). A two-stage melting proceeds via an intermediate phase whose nature depends on which order melts first: a) if $\theta_0$ disorders while $\vec{\eta}$ remains ordered, the intermediate phase is nematic that breaks $\mathcal{C}_3$ but preserves translation; b) if $\vec{\eta}$ disorders first but $\theta_0$ retains $Z_3$ phase coherence, we call the intermediate phase ``super"-fluid due to the phase coherence. Since $\cos(3\theta_0)$ is relevant at the would-be BKT transition of a true $\tilde N$ superfluid \cite{PhysRevB.16.1217}, we expect that thermal transition of $\theta_0$ to be three-state Potts-like. Therefore, in either route, both transitions, between SN and the intermediate phase, and between the intermediate and the fully disordered phase, are in the three-state Potts universality class, reflecting $Z_3$ symmetry breaking at each step.

\section{Other fillings}
\label{sec:other_filling}
\para Extending beyond $\nu=2/3$, the cluster constraint $\sum_{i\in\varhexagon_r} n_i\in\{0,1,2,3,4,5,6\}$ yields the “commensurate” fillings $\nu=0,1/3,2/3,1,4/3,5/3,2$. Since $\nu$ and $2-\nu$ are related by particle–hole symmetry and $\nu=0$ is trivial, it suffices to analyze $\nu=1/3$, $2/3$ and $1$. At $\nu=1/3$, Thm.~~\ref{thm:CL} applies directly (no need for the tile set $\Sigma_2$), so $\tilde{N}$ remains an exact symmetry of the constrained Hilbert space. In contrast, at $\nu=1$ the leading ring exchange breaks $\tilde N$ (Fig.~~\ref{fig:other-filling}a). A constructive scheme (see SM Sec. \ref{sec:app:ops}) can be proofed to generate all local operators order by order, yielding the leading ring exchanges (Fig.~~\ref{fig:other-filling}b) to be $\mathcal O_1$ for $\nu=1$, $\mathcal O_2$ for $\nu=2/3$, and $\mathcal L_2$ (the ``Lemniscate operator" in Ref.\cite{mao2023fractionalization}) for $\nu=1/3$. Notably, $\mathcal L_2$ conserves a sublattice dipole $\tilde{\vec P}=\sum_i n_A(x_i)\vec x_i-\sum_j n_A(x_j)\vec x_j$, which is violated by higher-order terms such as $\mathcal L'$. Therefore we also expect the SN phase to exist at $\nu=1/3$. For $\nu=1$, the $\mathcal O_1$ term gaps the system into a sublattice charge-ordered state.

\section{Summary and discussion}
\label{sec:sum}
\para In summary, we uncovered a robust mechanism by which geometric frustration and local Hilbert-space constraints enforce an emergent, globally conserved quantum number for bosons under cluster charging interaction on the honeycomb lattice. By mapping the constrained Hilbert space at $\frac{2}{3}$ ($\frac{4}{3}$) or $\frac{1}{3}$ ($\frac{5}{3}$) filling to a nontrivial tiling problem and invoking the Conway–Lagarias boundary invariant, we rigorously established global conservation of the sublattice polarization $\tilde{N}=N_A-N_B$, beyond the previously shown $U(1)\times U(1)$ gauge symmetry of classical ground state\cite{zhang2024bionicfractionalizationtrimermodel}. By mapping the local constraints to coupled double-dimers, we derive an effective field theory that faithfully incorporates the conservation of $\tilde{N}$ and the local constraints of the bosonic quantum Hamiltonian. Quantum fluctuations within this manifold give rise to an insulating yet macroscopically phase-coherent and gapless “supernematic” (SN) phase, characterized by a unique intertwined order that spontaneously breaks both the emergent global $U(1)$ symmetry and the lattice $C_3$ rotation, but not translation. We derived a \emph{non-local} order parameter representing the SN that can be detected through ODLOR: the ``bumped loop'' operator, which is markedly different from familiar Wilson loop through its shape-dependent charge under $\tilde{N}$. We also analyzed the finite temperature transitions that sequentially restore the broken symmetries. 

\para The proposed SN phase is unusual in many ways as a phase of matter. (i) It is a gapless insulator. (ii) It is a macroscopically phase-coherent state that is not a superfluid.  (iii) It is an intertwined order with $C_3$ breaking (nematic) that originates from the elongated resonance operators. (iv) It has a non-local order parameter that is distinct from the more familiar Wilson loop for its charge being shape-dependent. Naturally, it will be most exciting to realize SN in physical systems. Although our field-theoretic discussion only holds for bosonic systems, the conservation of $\tilde{N}$ due to Conway-Lagarias invariant at filling $1/3$ and $2/3$ of the quantum cluster-charging model is independent of statistics. Since the charged object under $\tilde{N}$ responsible for the condensation is bosonic in both bosonic and fermionic models, we expect that the SN could still be present in a fermionic trimer model, which we leave for future studies.

We discover the SN phase from the field theory perspective based on expanding the cosine term around its minima. While this expansion is valid deep in the SN phase, its validity near the phase transition to the CO phase is an open question that requires further study. Nevertheless, the validity of the SN phase as a possible ground state can be further supported by considering a distinct microscopic route, without tuning quantum fluctuations starting from the CO phase. A concrete realization is provided by a trimer generalization of the Rokhsar–Kivelson Hamiltonian, which we refer to as the trimer–Rokhsar–Kivelson (trimer-RK) Hamiltonian. This model naturally extends the RK construction for quantum dimer models~\cite{RK} and includes both a ring-exchange term, $\mathcal{O}_2$, and a potential term (see Appendix~\ref{sec:app:sq} for explicit definitions). At the RK point, where the strengths of these two terms are equal, the ground state is an equal-weight superposition of all trimer-covering configurations, analogous to the dimer model at its RK point. The equal-time correlation functions at the trimer-RK point is the same as the ones in trimer liquid \cite{zhang2024bionicfractionalizationtrimermodel}. In dimer models on bipartite lattices, such as the honeycomb lattice~\cite{moessner2001phase}, perturbations away from the RK point toward stronger exchange terms drive the system into an ordered phase. In contrast, quantum fluctuations around the trimer-RK point generate an effective action that stabilizes the SN phase (see Appendix~\ref{sec:app:sq}). The key distinction between the proposed trimer-RK and dimer-RK Hamiltonians is rooted in the $\tilde{N}$ conservation. In the trimer case, the dynamics including stochastic fluctuations must preserve $\tilde{N}$ exactly, leading to a Langevin equation of the same form as that describing volume-conserving surface growth~\cite{Sun1989PhysRevA.40.6763}. Upon quantization, this yields the phase-angle Lagrangian identical to Eq.~\ref{eq:eff_theta}, in contrast to the quantum Lifshitz model that governs the dimer RK point on bipartite lattices~\cite{Henley2004J.Phys.:Condens.Matterc}.

\para As a physical platform to persue the realization of SN, we propose the Rydberg atom array. The Rydberg blockade mechanism has been shown to provide cluster-charging constraints on honeycomb lattices \cite{kornjavca2023trimer}. The algebraic decay of various operators of the SN phase can be directly measured in this platform. Measuring the non-local order parameter can be challenging, but it may be achievable through strategies analogous to that used for topological phases in Ref.~\cite{Semeghini2021Sciencea}. Since the total number of particles is not fixed in Rydberg systems, we need to consider the stability of the SN under doping away from the commensurate fillings ($2/3$ or $1/3$). As was discussed in the thermal violation of $\tilde{N}$ conservation, the monomers induce a threefold anisotropy, which pins the macroscopic phase angle of the SN.  Doping also introduces monomers that carry fractionalized charges of $e/3$~\cite{mao2023fractionalization, zhang2024bionicfractionalizationtrimermodel}.  However, since the $\vec{\eta}$ field is pinned in the SN phase, we expect these monomers to be confined at low temperature and to become deconfined only in the ``super"-fluid regime at finite temperature.

\para Our unprecedented use of tiling invariants to discover a new quantum many-body state opens a new path for quantum-state preparation to approach a classically hard problem: the tiling combinatorial optimization problem. 
While we have focused on a specific set of tiles, it would be interesting to explore whether the notion of a boundary invariant extends to other tilings or to other lattices.
An open question is whether the operator $\mathcal{O}_2$ is ergodic in the Hilbert space of configurations satisfying local constraints, or whether geometrical constraints generate hidden conserved quantities beyond $\tilde{N}$ and the 1-form symmetries.
Physical realization of coherent superpositions of trimer configurations would provide a powerful means to probe such questions, offering direct access to the structure of constrained Hilbert spaces where ergodicity may break down.
More broadly, these simulations could shed light on the quantum dynamics underlying hard combinatorial optimization problems, where quantum interference could provide a speedup over classical methods.

\para {\it Acknowledgement.} -- We thank Jiaxin Qiao, Yizhi You, Cenke Xu, Yi-Zhuang You, Jing-Yuan Chen, Zheng-Cheng Gu, T. Senthil, Matthew Fisher, and Roderig Moessner for the stimulating discussions. We thank Christopher Mudry and Eduardo Fradkin for feedbacks on the draft. DM is supported by SNSF Grant No. 219339. E-AK was supported in part by the NSF through the grant OAC-2118310 supported by the U.S. Department of Energy through Award Number: DE-SC0023905.
This research is funded in part by the Gordon and Betty Moore Foundation’s EPiQS Initiative, Grant GBMF10436 to E-AK and grant NSF PHY-2309135 to the Kavli Institute for Theoretical Physics (KITP).

\bibliographystyle{apsrev4-2}
\bibliography{refs}

\appendix

\section{Details on symmetry properties of the height fields}
\label{sec:app:sym}
\subsection{Colomnar ordering states}
\begin{figure*}
    \centering
    \includegraphics[width=0.95\linewidth]{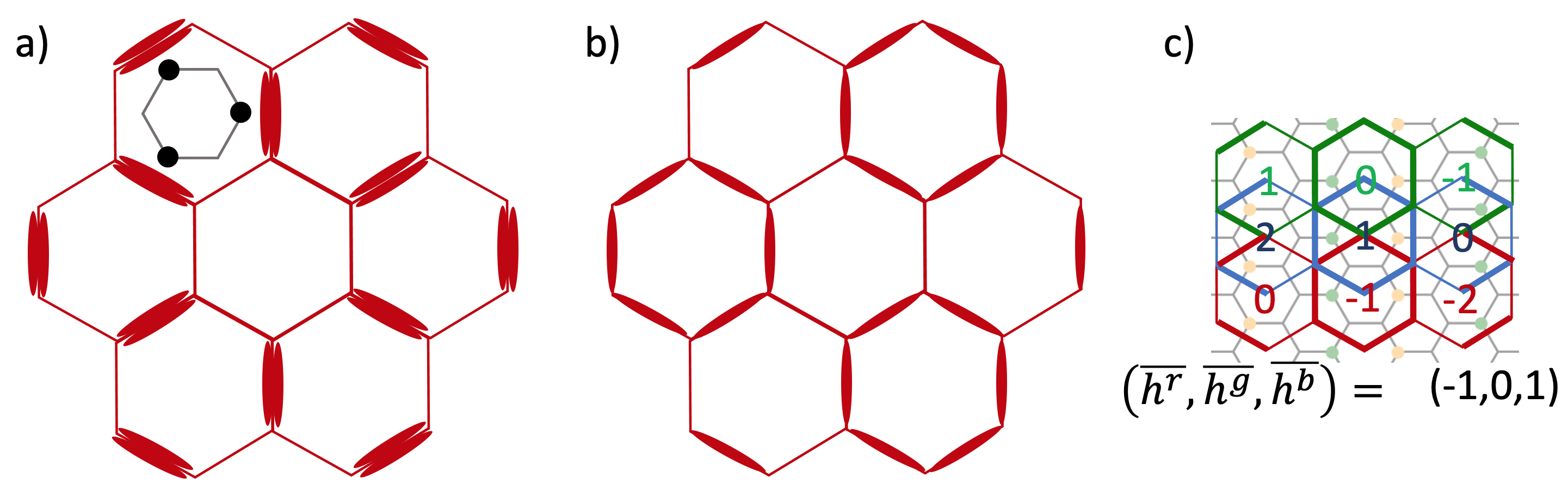}
    \caption{Two possible types columnar orders in the honeycomb lattice double-dimer model. a) With double occupation, the corresponding physical configuration has three particles of some plaquettes, violating cluster-charging constraints at filling $2/3$. One of such plaquette configurations is labelled by the black dots. b) Columnar orders with loops around the honeycomb plaquettes. c) The colomnar order of the physical configuration (in the background) and the corresponding colomnar orders of the three double-dimer models (only the three consecutive plaquettes are shown), with the corresponding height fields. The height fields of each color has the three sublattice structure. Bottom : the average of the height fields of the three sublattices of each color.}
    \label{fig:app:colomnar}
\end{figure*}

For the dimer models on bipartite lattice, the ``flat" height field configurations of the coarse-grained heights are the columnar ordering. For the double-dimer model on the honeycomb lattice of our interest, there are two possible colomnar ordering: the ones with double occupation (Fig. \ref{fig:app:colomnar} a)) Eand the ones with loops (Fig. \ref{fig:app:colomnar} b)). However, when mapping back to physical particle configurations, the colomnar ordering with double occupation violates the cluster-charging constraints at filling $2/3$ because some plaquettes are forced to host three particles. Hence we will only consider the colomnar ordering with loops in the following.

There are three distinct columnar orderings related by lattice translation for one color. Naively, there are $3^3 = 27$ columnar orderings taking into account the three colors. However, there are only $9$ of them that can be mapped to physical configurations, one of which listed in Fig. \ref{fig:app:colomnar} c) with the height fields. The other COs can be related to the one in Fig. \ref{fig:app:colomnar} c) by lattice translation and rotation.

\subsection{Symmetry properties under lattice transformations}
Microscopically, the height fields transform under the lattice symmetries as the following,
\begin{equation}
\begin{split}
        &\mathcal{T}_1:~h^{r,g,b}(x) \rightarrow h^{g,b,r}(T_1 x) \\
        &\mathcal{T}_2:~h^{r,g,b}(x) \rightarrow h^{b,g,r}(T_2 x) \\
        &\mathcal{C}_3:~h^{r,g,b}(x) \rightarrow h^{r,g,b}(C_3 x)\\
        &\mathcal{I}:~h^{r,g,b}(x) \rightarrow - h^{r,b,g}(-x),
    \end{split}
\end{equation}
where the center of the $C_3$ rotation and inversion is taken to be around the center of a red plaquette.

\begin{figure*}
    \centering
    \includegraphics[width=0.85\linewidth]{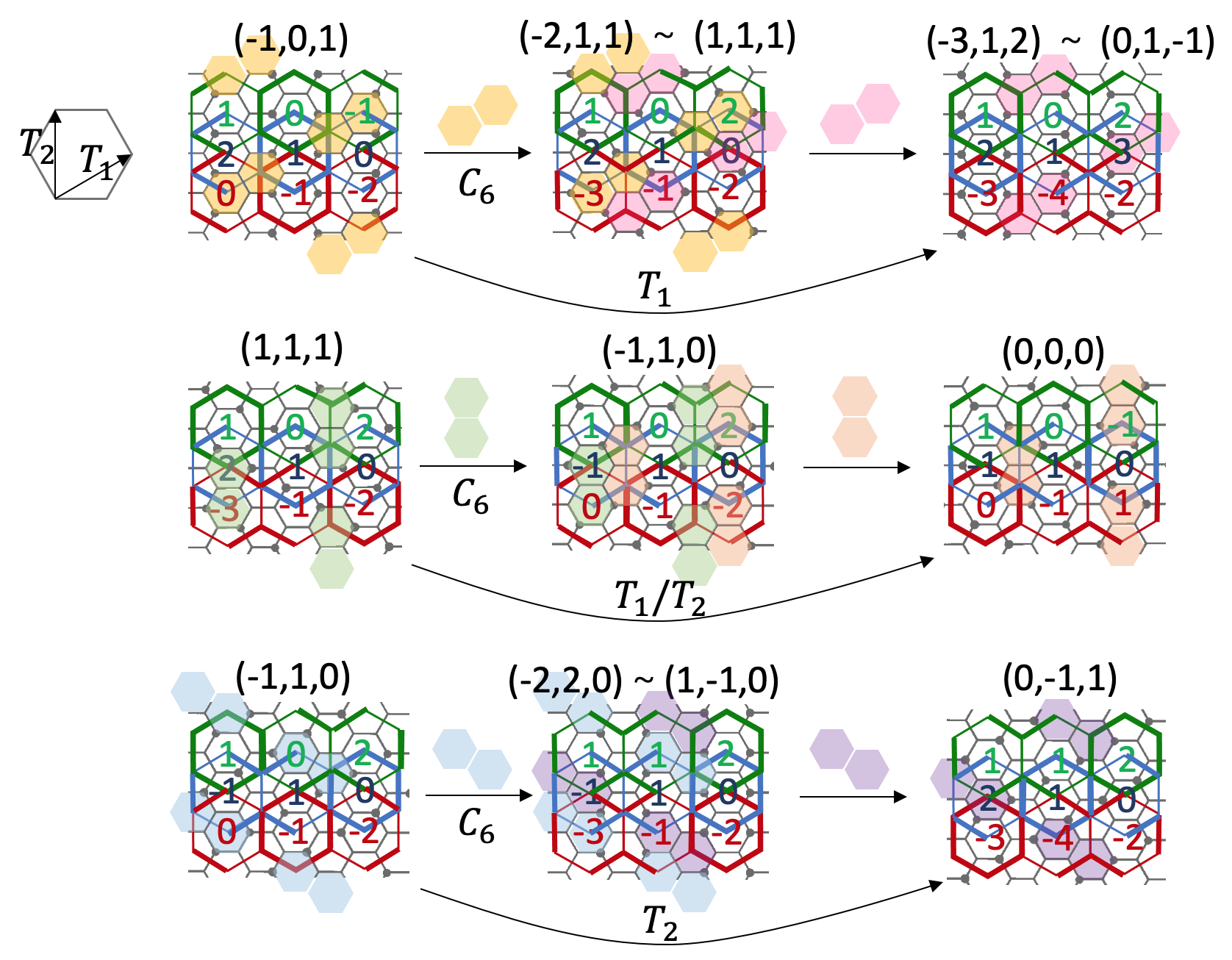}
    \caption{Change of height under applying local operators. Upperleft: definition of lattice translations. For each CO, the three numbers above denote the $(\tilde{h}^r, \tilde{h}^g,\tilde{h}^b)$, taking average over the heights of the three sublattices of the corresponding color (colored numbers in the center of the plaquettes). Note that we only present the dimer configuration and heights of three plaquettes but the pattern is repeated periodically. The shaded double plaquette denotes where a $\mathcal{O}_2$ (or $\mathcal{O}_2^\dag$) is applied, which matches the symbols above the arrows, denoting the ``before" and ``after" applying of the operator. The corresponding lattice transformation is denoted underneath the arrow. We also note that each $(\tilde{h}^r, \tilde{h}^g,\tilde{h}^b)$ is defined mod $3$. Here we only present seven out of nine COs but the transformation of the rest can be worked out in the similar fashion.}
    \label{fig:height-trans}
\end{figure*}

In the continuum, the coarse-graining introduces non-zero average values of $h^{r,g,b}$, and the lattice symmetry can act with some additional shift on top of the microscopic translation \cite{Fradkin2004Phys.Rev.Bb}. To see this clearly, we consider the nine colomnar ordering states (flat configurations with constant average colored height fields). The lattice transformation among these COs can be implemented via applying local operators across the lattice, allowing us to derive the values of height fields starting from one specific CO (see Fig.\ref{fig:height-trans}). Hence, under lattice symmetries, the average heights transform as such,
\begin{equation}
    \begin{split}
        &\mathcal{T}_1:~\tilde{h}_0 \rightarrow \tilde{h}_0- 1;~ \tilde{\eta} \rightarrow \tilde{\eta} e^{-i \frac{4\pi}{3}}\\
        &\mathcal{T}_2:~\tilde{h}_0 \rightarrow \tilde{h}_0 - 1;~ \tilde{\eta} \rightarrow \tilde{\eta} e^{i \frac{4\pi}{3}}\\
        &\mathcal{C}_3:~\tilde{h}_0 \rightarrow \tilde{h}_0; ~ \tilde{\eta} \rightarrow \tilde{\eta} - i \frac{\sqrt{3}}{3}\\
        &\mathcal{I}:~ \tilde{h}_0  \rightarrow - \tilde{h}_0 ;~  \tilde{\eta} \rightarrow -  \tilde{\eta},
    \end{split}
\end{equation}
where
\begin{equation}
\begin{split}
        \tilde{h}^r &= \tilde{h}_0  + \tilde{\eta} + \tilde{\eta}^*\\
       \tilde{ h}^g &= \tilde{h}_0  + \tilde{\eta} e^{-i \frac{4\pi}{3}} + \tilde{\eta}^*e^{i \frac{4\pi}{3}}\\
        \tilde{h}^b &= \tilde{h}_0  + \tilde{\eta} e^{i \frac{4\pi}{3}} + \tilde{\eta}^*e^{-i \frac{4\pi}{3}}.
\end{split}
\end{equation}
Here we use $\tilde{.}$ to denote the avearge value of the height fields around three sublattices of each color.

Adding up the shift in the average heights and the microscopic transformation gives us the transformation of the height fields in the continuum, namely, Eq.\ref{eq:trans:h}.

\subsection{Relating coarse-grained fields to microscopic fields}
\label{sec:app:h}
Recall Eq. \ref{eq:p_duality}, microscopically, 
\begin{equation}
    n_A = - \frac{1}{3}\sum_{\triangleright} h + \frac{1}{3},~ n_B =  \frac{1}{3}\sum_{\triangleleft} h + \frac{1}{3}.
    \label{eq:app:n}
\end{equation}
If we perform a gradient expansion of the above equations, we obtain $\partial \eta$ terms to the leading order,
\begin{equation}
    n_A - \frac13 \sim - e^{-i \frac{2\pi}{3}(2n_2 +n_1)} (\partial_x + i \partial_y) \eta +h.c.+ ...,
\end{equation}
where we denote the coordinate of $n_A$ as $n_1 \vec{a}_1 + n_2 \vec{a}_2 \equiv (x_A, y_A)$.  $\vec{a}_1 = (\frac{\sqrt{3}}{2}, \frac{1}{2})$ and $\vec{a}_2 = (0,1)$ are the lattice vectors. The prefactor has the three sublattice structure corresponding to the three different $r,g,b$ height fields positions around a given A site. The similar contribution in $n_B$ can be written as,
\begin{equation}
    n_B - \frac13 \sim e^{-i \frac{2\pi}{3}(2n_2 +n_1)} (\partial_x - i \partial_y) \eta +h.c.+ ...,
\end{equation}

Note that in the gradient expansion, there is also a term that is propotional to $h_0$. Since $h_0$ is a compact field, we should find a vertex operator that has the same behavior around small $h_0$, which is the $\mathcal{V}_{(1,1,1)}$. This contributes to $n_A$ and $n_B$ as,
\begin{equation}
\begin{split}
n_A - \frac{1}{3} &\sim  - \frac{1}{2\pi} \sin{2\pi h_0}\\
n_B - \frac{1}{3} &\sim  \frac{1}{2\pi} \sin{2\pi h_0}
\end{split}
\label{eq:nh0}
\end{equation}

However, this is not the full story due to the smearing in the coarse-graining. To the leading order in vertex operator, this amounts to match the lattice configuration of the colomnar ordering and the corresponding uniform heights in the continuum, analogous to dimer model \cite{Fradkin2004Phys.Rev.Bb}, which yields,
\begin{equation}
\begin{split}
    &n_A - \frac{1}{3} \\
    \sim& \frac{1}{9}\Big[ e^{i \frac{2\pi}{3}(\frac{x_A}{\sqrt{3}} + y_A) - i \frac{2\pi}{3}}(e^{i \frac{2\pi}{3} h^r} + e^{i \frac{2\pi}{3} h^g}+ e^{i \frac{2\pi}{3} h^b})\Big]\\
    +& \frac{1}{9}\Big[ e^{-i \frac{2\pi}{3}\frac{2x_A}{\sqrt{3}} - i \frac{2\pi}{3}}(e^{i \frac{2\pi}{3} h^r} + e^{i \frac{2\pi}{3} h^g - i \frac{2\pi}{3}}+ e^{i \frac{2\pi}{3} h^b + i \frac{2\pi}{3}})\Big]\\
    +& \frac{1}{9}\Big[ e^{-i \frac{2\pi}{3} (-\frac{x_A}{\sqrt{3}} + y_A) - i \frac{2\pi}{3}}(e^{i \frac{2\pi}{3} h^r+ i \frac{2\pi}{3}} + e^{i \frac{2\pi}{3} h^g- i \frac{2\pi}{3}}+ e^{i \frac{2\pi}{3} h^b })\Big]\\&+h.c.,
\end{split}
\end{equation}
where $h^{r,g,b}$ have the corresponding coordinates according to the microscopic relation Eq.\ref{eq:app:n}.

Plugging in the pinned values for $\vec{\eta}$, we have,
\begin{equation}
\begin{split}
    &n_A - \frac{1}{3} \sim \frac{1}{9}e^{i \frac{2\pi}{3}h_0(x_A)}\sum_{\triangleright}  f_\alpha (x_A) + h.c., 
\end{split}
\label{eq:nh0G}
\end{equation}
where the summation is taken over the $r,g,b$ heights around the site $x_A$, denoted by $\alpha$, and the factors $f_{\alpha}(x_A)$ can be written as,
\begin{equation}
    \begin{split}
        &f_r = e^{\frac{i 2\pi e_1}{3}} \left(e^{i G_1 \cdot r_A - i \frac{2\pi}{3}} + e^{i G_2 \cdot r_A - i \frac{2\pi}{3}} + e^{i G_3 \cdot r_A }\right)\\
        &f_g = e^{\frac{i 2\pi e_2}{3}}\left(e^{i G_1 \cdot r_A - i \frac{2\pi}{3}} + e^{i G_2 \cdot r_A + i \frac{2\pi}{3}} + e^{i G_3 \cdot r_A + i \frac{2\pi}{3}}\right)\\
        &f_b = e^{\frac{-i 2\pi (e_1+e_2)}{3}}\left(e^{i G_1 \cdot r_A - i \frac{2\pi}{3}} + e^{i G_2 \cdot r_A } + e^{i G_3 \cdot r_A - i \frac{2\pi}{3}}\right),
    \end{split}
    \label{eq:fG}
\end{equation}
where $G_1 = \frac{2\pi}{3}(\frac{1}{\sqrt{3}}, 1)$, $G_2 = \frac{2\pi}{3}(-\frac{2}{\sqrt{3}}, 0)$, and $G_3 = \frac{2\pi}{3}(\frac{1}{\sqrt{3}}, -1)$.

Similarly, for sublattice B,
\begin{equation}
\begin{split}
    &n_B - \frac{1}{3} \\
    \sim& \frac{1}{9}\Big[ e^{i \frac{2\pi}{3}(\frac{x_B}{\sqrt{3}} + y_B) + i \frac{2\pi}{9}}(e^{i \frac{2\pi}{3} h^r} + e^{i \frac{2\pi}{3} h^g}+ e^{i \frac{2\pi}{3} h^b})\Big]\\
    +& \frac{1}{9}\Big[ e^{-i \frac{2\pi}{3}\frac{2x_B}{\sqrt{3}} + i \frac{2\pi}{9}}(e^{i \frac{2\pi}{3} h^r} + e^{i \frac{2\pi}{3} h^g - i \frac{2\pi}{3}}+ e^{i \frac{2\pi}{3} h^b + i \frac{2\pi}{3}})\Big]\\
    +& \frac{1}{9}\Big[ e^{-i \frac{2\pi}{3} (-\frac{x_B}{\sqrt{3}} + y_B) + i \frac{2\pi}{9}}(e^{i \frac{2\pi}{3} h^r+ i \frac{2\pi}{3}} + e^{i \frac{2\pi}{3} h^g- i \frac{2\pi}{3}}+ e^{i \frac{2\pi}{3} h^b })\Big]\\&+h.c..
\end{split}
\end{equation}
where the origin of the coordinate systems is set to be at A site. We also have,
\begin{equation}
\begin{split}
    &n_B - \frac{1}{3} \sim \frac{1}{9}e^{\frac{i 8\pi}{9}} e^{i \frac{2\pi}{3}h_0(x_B)}\sum_{\triangleleft}  f_\alpha (x_B) + h.c..
\end{split}
\label{eq:nh0GB}
\end{equation}

To summarize, in terms of the coarse-grained fields, to the leading orders, there are three contributions to the density $n_{A,B}$: a $\partial \eta$ term (Eq. \ref{eq:app:n}), a $\sin{2\pi h_0}$ term (Eq. \ref{eq:nh0}) and a $e^{i \frac{2\pi}{3} h_0}$ term with spatially periodic coefficients (Eq. \ref{eq:nh0G}).

\section{Sublattice density structure factor}
\label{sec:sf}
Before integrating out the $h_0$ field in the effective action of the phase angle $\theta_0$ in Eq.\ref{eq:eff_theta}, we have
\begin{equation}
    \mathcal{S}_1[h_0, \theta_0] = \int -i h_0 \partial_\tau \theta_0 + \frac{\tilde{\mathbf{y}}}{4} h_0^2 + \tilde{t}(\partial_\mu \theta_0 )^2+ ....
\end{equation}
From the equation of motion of $h_0$,
\begin{equation}
    h_0 = \frac{i 2 \partial_\tau \theta_0}{\tilde{\mathbf{y}}}.
\end{equation}
Hence, in Matsubara frequency, the correlation function of $h_0$ in momentum space is,
\begin{equation}
    \chi_{h_0} (q, i\omega_n) = \frac{4}{\tilde{\mathbf{y}}^2}\omega_n^2 \chi_{\theta_0} = \frac{2}{\tilde{\mathbf{y}}}\frac{\omega_n^2}{\omega_n^2 + v^2 q^2},
\end{equation}
where $v = \sqrt{\tilde{t}\tilde{\mathbf{y}}}$ is the velocity of the Goldstone mode.

Analytic continuation $i \omega_n \rightarrow \omega + i 0^+$ of $\chi_{h_0}$ gives the retarded Green's function $\chi_{h_0}^R (q, \omega)$. The dynamical structure factor $S(q,\omega)$ follows from the fluctuation-dissipation theorem. At zero temperature,
\begin{equation}
    S(q,\omega) = \frac{-\Theta(\omega)}{\pi} \Im{\chi_{h_0}^R (q, \omega)} = \frac{v|q|}{\tilde{\mathbf{y}}}\delta(\omega - v |q|),
\end{equation}
and the static structure factor $S(q) = \int d\omega S(q,\omega)$.

Now let us consider the structure factor of the sublattice density $\tilde{n}(q) \equiv \sum_i n_{i,A} e^{i q\cdot x_{i,A}} - \sum_j n_{j,B} e^{i q \cdot x_{j,B}}$. From Eq.\ref{eq:nh0}, \ref{eq:nh0G} and \ref{eq:nh0GB}, we can write $\tilde{n}(q)$ in terms of $h_0$ fields. It is readily seen that 
\begin{equation}
\begin{split}
\tilde{S}(q,\omega) \sim& A \frac{v|q|}{\tilde{\mathbf{y}}}\delta(\omega - v |q|) \\
&+ B \sum_{i} \frac{v|q + G_i|}{\tilde{\mathbf{y}}}\delta(\omega - v |q+G_i|)  ~ \text{for q $\rightarrow$ 0},
\end{split}
\end{equation}
where $G_{i= 1,2,3} = G_{1,2,3}$ defined in Eq.\ref{eq:fG} and $G_{i=4,5,6} = -G_{1,2,3}$, and $A,B$ are some constants. In the above equation, the first term comes from correlation between $\sin(2\pi h_0)$ and the second term from the ones between $e^{i \frac{2\pi}{3} h_0} f_\alpha(x)$.

\section{Systematic construction of local operators}
\label{sec:app:ops}
\begin{figure}
    \centering
    \includegraphics[width=0.5\linewidth]{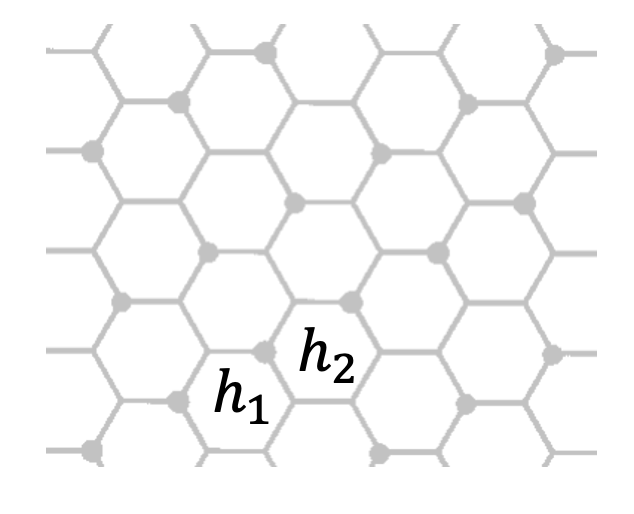}
    \caption{Gauge choice for the height fields.}
    \label{fig:choice}
\end{figure}

Here we give a systematic construction of a set of local operators that commutes with the cluster charging constraints, denoted as $\mathbb{S}_{op}$. We then prove that the set of nonzero local operators at a particular filling fraction, denoted as $\mathbb{S}_{op-nz}$, is always a subset of $\mathbb{S}_{op}$.
% \begin{figure*}
%     \centering
%     \includegraphics[width=0.8\linewidth]{ref_state.png}
%     \caption{Examples of reference states of a) filling 1 and b) filling 1/3, where the numbers in the plaquette centers denote $(N_0, N_0')$. }
%     \label{fig:ref_state}
% \end{figure*}
To construct the set of local operators, we first consider how to represent the configurations in a manner that is convenient for comparing the difference between them. It turns out that the relationship between the particle number and the height field (Eq.\ref{eq:p_duality}) can be generalized to filling $1$ and $1/3$,
\begin{equation}
    n_A = - \frac{1}{3}\sum_{\triangleright} h + \frac{\nu}{2},~ n_B =  \frac{1}{3}\sum_{\triangleleft} h + \frac{\nu}{2}.
    \label{eq:height_general}
\end{equation}
For a given particle configuration $n_i$'s, the mapping to height fields is not one-to-one, namely, gauge redundency in Eq.\ref{eq:height_general}. We can fix the gauge by fixing the value of the height fields corresponding to two different colors. For instance, we can choose to fix $h_1$ and $h_2$ to be constant integers in Fig. \ref{fig:choice}. After the gauge-fixing, the mapping between the $n_i$ configuration and the height field is injective, using Eq.\ref{eq:height_general}, and the cluster-charging condition is satisfied by construction. Therefore, it is readily seen that the following lemma hold,

\begin{lemma}
    For two configurations with the same $\nu$, satisfying cluster-charging condition, differ locally, their height field representations also differ locally after gauge fixing.
  \label{lemma1}
\end{lemma}

Therefore, we can construct all the possible local operators for $n_i$'s if we can find a way to construct local operators for $h_j$'s. We denote the set of local operators in the height representation as $\mathbb{S}_{op}$. The construction is in the following subsection.

However, the mapping between particle number configuration and the height configuration is not surjective. Since Eq.\ref{eq:height_general} only ensure that integer-valued $n_i$'s maps to integer-valued $h_j$'s, given fixed integer $h_{1,2}$, it does not exclude the possibility that some configuration of $h_j$'s can map back to ``unphysical" $n_i$'s, such as non-integer values or negative values, despite satisfying the cluster-charging condition. In particular, there are operators which we call ``zero operators" that are local in the height representation but annihilate all the physical particle configurations. We will discuss those zero operators in the subsection after the next. Hence the set $\mathbb{S}_{op}$ is strictly larger than the set of the physical local operators acting on $n_i$'s, denoted as $\mathbb{S}_{op-nz}$, which are the subset containing ``non-zero" operators. 

\subsection{Local operators in $\mathbb{S}_{op}$}
Now let us consider the effect of local operators on the height fields defined in the previous section. From Lemma \ref{lemma1}, the local operator in $n_i$'s changes $h_j$ fields locally, which we denote as $\Delta h$. Due to the relationship Eq.\ref{eq:height_general}, the minimal change in the height is $\Delta h = \pm 3$.

\begin{figure}
    \centering
    \includegraphics[width=\linewidth]{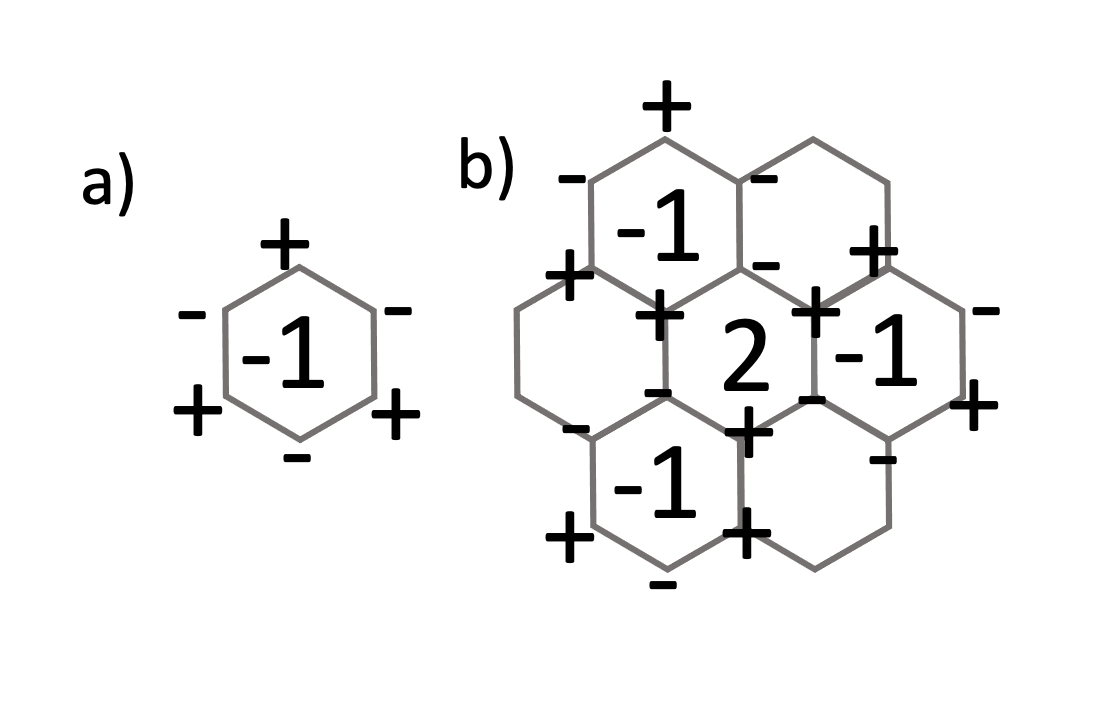}
    \caption{Local operators constructed from $\Delta h/3$.}
    \label{fig:deltaN}
\end{figure}

Let us consider $\Delta h_r = -3$ for plaquette $r$. By definition, it means that around plaquette $r$, the occupations on $A$ sublattice decrease by 3 and those on $B$ sublattice increase by 3, which is precisely the action of operator $\mathcal{O}_1$ (Fig. \ref{fig:deltaN} a)).

Since $\Delta h/3$ can be any integer, the operator can also have ``strength" other than $1$. This does not make sense by itself since we cannot annihilate/create a particle more than once due to hard core constraints. However, such $\Delta h/3$ pattern can still make sense after taking into account the nearby plaquettes as long as the change in the resulting $n_i$ fields is still $\pm 1$ or $0$. See Fig.\ref{fig:deltaN} b) for an example. 

Since all the local operators can be viewed as local patterns of $\Delta h/3$ fields and all the $\Delta h/3$ patterns can be viewed as stacking and combining the $\mathcal{O}_1$'s, we obtain a systematic way of constructing all the possible local operators.

\subsection{Zero operators}
In the previous section, we derive a systematic construction of the set $\mathbb{S}_{op}$. Now we will show that there are the so-called ``zero operators" in $\mathbb{S}_{op}$, which annihilate all the states in the constrained physical Hilbert space at a certain filling $\nu$.

\begin{figure*}
    \centering
    \includegraphics[width=0.8\linewidth]{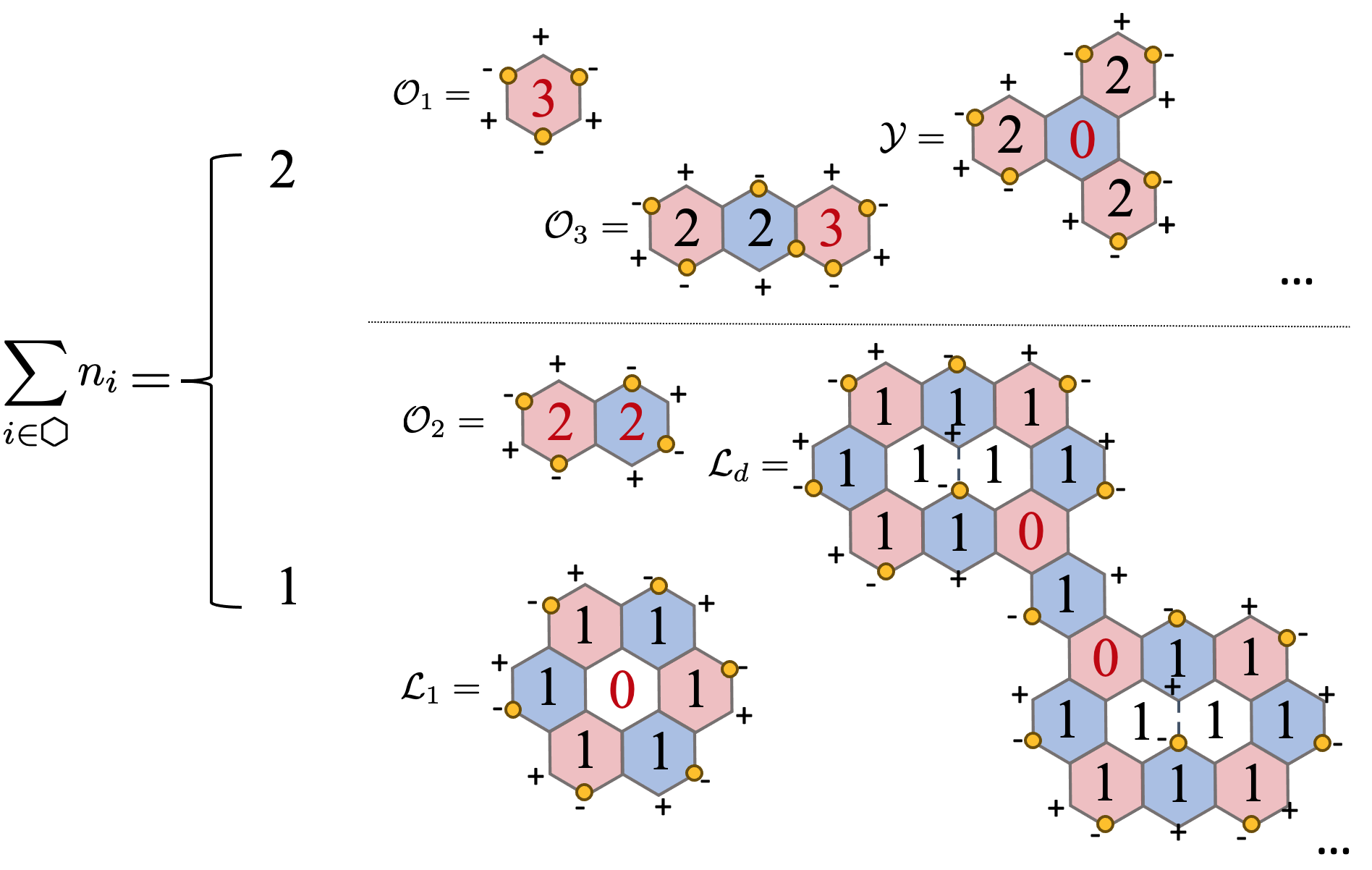}
    \caption{Examples of zero operators for filling 2/3 and 1/3. The color pink and blue denote $\mathcal{O}_1$ with strength $\pm 1$. The numbers on the plaquette denote the sum of the particle numbers around it, which enables the corresponding operator to act non-vanishingly. In particular, the red number denotes the contradiction with the filling constraints.}
    \label{fig:zero_ops}
\end{figure*}

Here we give some examples of zero operators (Fig.\ref{fig:zero_ops}). The obvious type of zero operators is that it has more $+$ (or $-$) signs around a plaquette than what the filling fraction imposes, such as $\mathcal{O}_1$ for filling 2/3 and $\mathcal{O}_2$ for filling 1/3. However, there are more subtle zero operators, such as $\mathcal{O}_3$ and $\mathcal{Y}$ for filling 2/3 and $\mathcal{L}_{1,d}$ for filling 1/3. So far, we are only able to tell that these are zero operators by considering the particle configurations upon which they act non-vanishingly and finding a contradiction to the imposed filing fraction.

\section{Tile mapping and quantum number for filling $1/3$}
\label{sec:one-third}
\begin{figure}
\centering
    \includegraphics[width=0.9\linewidth]{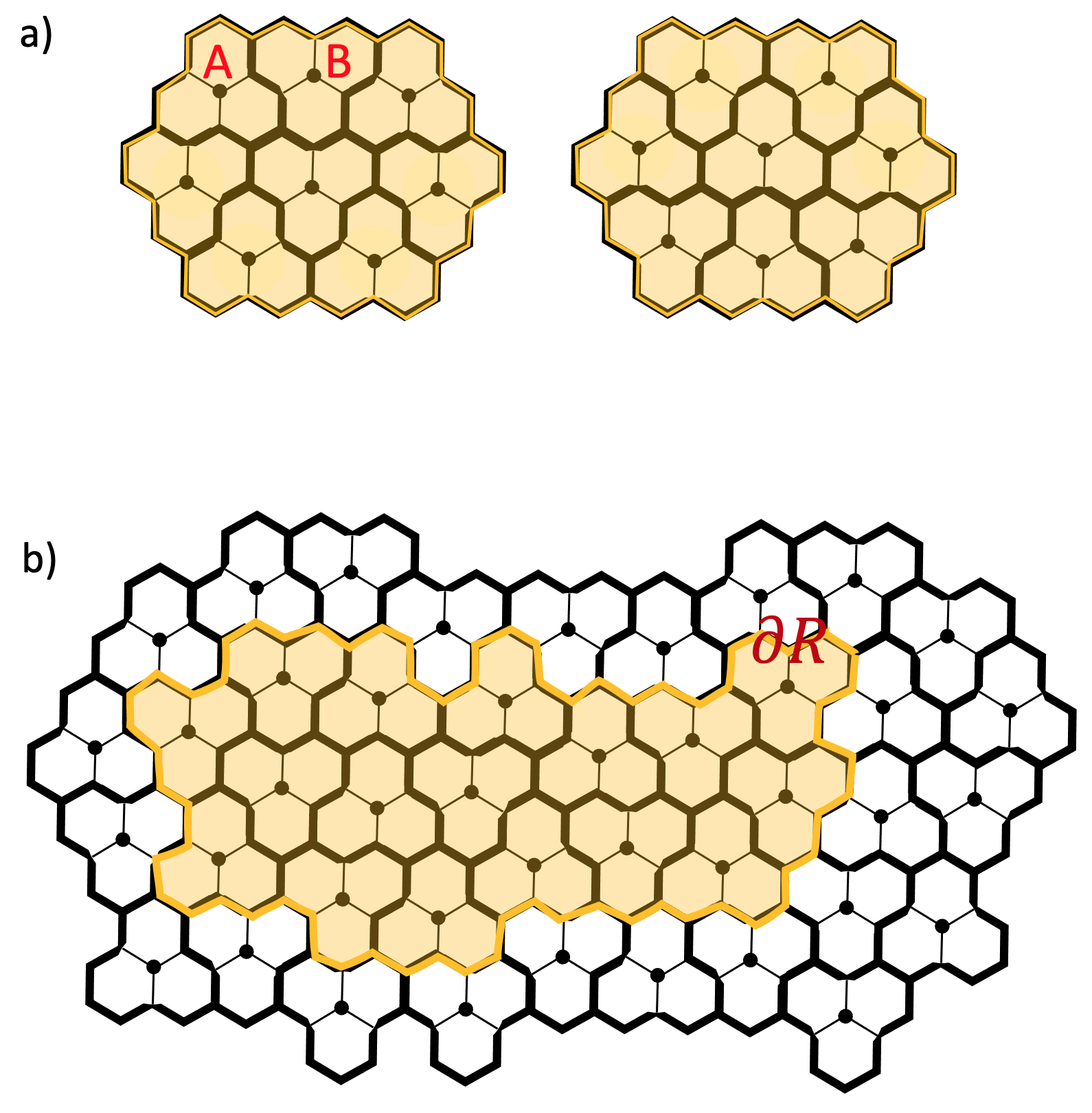}
    \caption{a) Mapping between the particles on A and B sublattices (black dots) and the tiles (three cojoining hexagonal plaquettes) and the corresponding tiling pattern for the before and after states upon the action of the lemniscate operator $\mathcal{L}_2$. The orange shades denote the area within which the tiles rearrange. b) An example of a region $R$ (orange shade), with boundary $\partial R$ (orange contour). }
    \label{fig:filling13}
\end{figure}
For filling 1/3, there is a one-to-one correspondence between a charge configuration and a tiling pattern of a triangular tiles on honeycomb lattice (Fig.\ref{fig:filling13} a) ), where the particles on the sites of the honeycomb lattice are denoted as black dots in the center of the tiles. The particle on the $A$ ($B$) sublattice of the honeycomb lattice is mapped to tile $T_1$ ($T_2$) and the cluster charging constraint is satisfied by the tiling conditions that there is no gap or overlap allowed between the tiles. Therefore, applying Thm. \ref{thm:CL}, we find that for any tilable region $R$ with a boundary $\partial R$, ( see Fig.\ref{fig:filling13} b) for an example of $R$), the difference of the sublattice particle number within $R$, i.e., $(N_A- N_B)_R \equiv \sum_{i \in R} (n_{i,A}- n_{i,B})$ only depends on the ``shape" of the boundary, i.e., the word generated by $\partial R$. Therefore via the same logic as in the filling $2/3$ case, $\tilde{N}$ is a conserved quantity under any action of local operators.

We note that the mapping from particles to tiles in the case of filling $1/3$ cannot be applied directly to filling $2/3$. The reason is that if we map the particle to tiles as in Fig.\ref{fig:filling13} a), we have overlapping tiles, i.e. each hexagonal plaquette gets touched twice by the tiles. For such overlapping tiling configurations, there is no direct generalization of Thm. \ref{thm:CL} since the overlapping tiling cannot always be separated as two regular tilings with no overlaps and no gaps. In the main text, we overcome this difficulty by creating another mapping that maps the constraints for filling $2/3$ to yet another nonoverlapping tiling problem, with edge matching puzzles as the building block.

\section{Perturbing around trimer-RK point}
\label{sec:app:sq}
\para In the quantum dimer model on bipartite lattice, the dimer constraints can be viewed as Gauss law on the sites of the lattices  \cite{Fradkin2004Phys.Rev.Bb}, and the low-energy physics can be described by compact $U(1)$ gauge theory in $2+1D$. The quantum fluctutation of the local moves in the dimer model corresponds to the instanton events and leads to confinement. Therefore, the gapless  RK point is in general unstable under quantum fluctuation. We can ask the analogous question for the trimer model, that is, starting from the trimer RK point where the wavefunction is an equal weight superposition of all possible classical trimer configuration, what is the role of quantum fluctuation? 

\begin{figure}[h]
    \centering
    \includegraphics[width=0.5\linewidth]{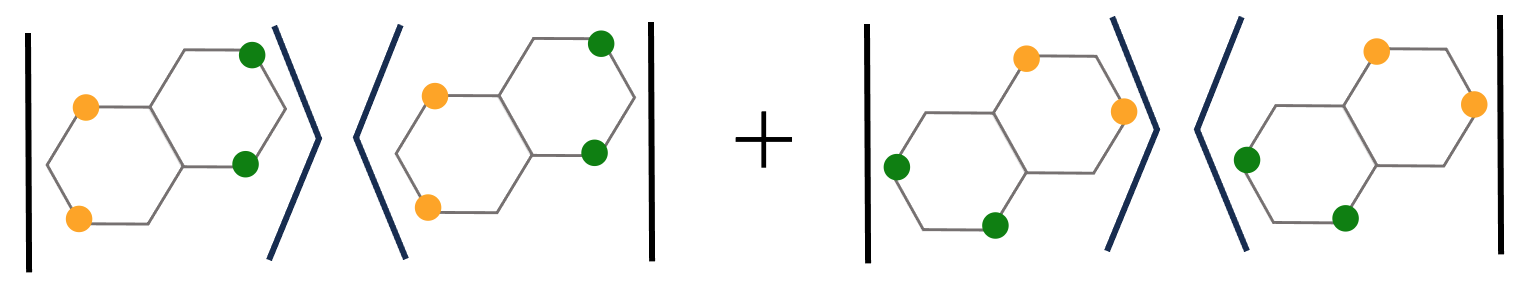}
    \caption{Potential term $\mathcal{P}$}
    \label{fig:P_V}
\end{figure}

\para For simplicity let us focus on filling $2/3$. The trimer-RK Hamiltonian can be written as,
\begin{equation}
    H_{\text{t-RK}} = - t \sum_{i,\alpha} (\mathcal{O}_{2}^\alpha(x_i)+ h.c.) + V \sum_{i,\alpha} \mathcal{P}^\alpha(x_i), 
\end{equation}
where $\mathcal{P}^\alpha(x_i)$ is the potential term and $\alpha$ denotes its orientation (see Fig.\ref{fig:P_V} for one orientation).
The RK point of the Hamiltonian $H_{\text{t-RK}}$ is when $t=V$, at which the ground state is an equal weight superposition of all possibile particle configurations at filling $2/3$, subject to the cluster-charging constraints. For each topological sector with fixed $\tilde{N}$ and fixed $U(1) \times U(1)$ winding numbers, there is one such state. 

Now we would like to derive the effective action, perturbing away from the RK point. We do so by first considering the height field representation. As pointed out in the main text, the $h^{R,G,B}$ fields can be recast into $h_0$ and $\vec{\eta}$ fields.
Classically, the entropy functional takes a Gaussian form of the fluctuation around the flat configuration. For dimer models on bipartite lattice, this yields $e^{- \int (\partial_\mu h)^2}$, since the dimer density is $\sim \partial h$. However, as noted in Sec. \ref{sec:app:h}, in the relationship between the density and the heights, $\vec{\eta}$ and $h_0$ enter differently. Roughly speaking, we have $n \sim e^{i K x} \partial \eta + \sin{h_0}$. Hence the entropy functional $S_{cl} = S_{cl}[\vec{\eta}]+ S_{cl}[h_0]$, where
\begin{equation}
S_{cl}[\vec{\eta}] = \frac{1}{2} K_\eta (\partial_\mu \vec{\eta})^2,
\end{equation}
and
\begin{equation}
S_{cl}[h_0] = m h_0^2.
\end{equation}
The coefficient $K_{\eta}$ can be obtained by matching the Gaussian action of the classical trimer model in Ref. \cite{zhang2024bionicfractionalizationtrimermodel}.
\para In the dimer model near RK point, it was realized by Henley \cite{Henley1997JStatPhysa, Henley2004J.Phys.:Condens.Matterc} that in the long wavelength limit, the plaquette flip dynamics can be captured by a Langevin equation of the height field (similar technique is also under the name of ``stochastic quantization", see \cite{parisi1981perturbation, DIJKGRAAF2010365} and the reference therein). Let us only focus on the $h_0$ field. The dynamics of $h_0$ follows the following Langevin equation,
\begin{equation}
    \partial_\tau h_0(\vec{r},\tau) = -f[h_0] + \zeta(\vec{r},\tau),
    \label{eq:langevin}
\end{equation}
where $f[h_0] = \frac{\delta S_{cl}[h_0]}{\delta h_0} = 2 m h_0$. $\zeta(\vec{r},\tau)$ is a noise term, which we will specify later.

\para In dimer model with plaquette flipping dynamics, $\zeta(\vec{r},\tau)$ is taken to be a Gaussian white-noise. However, this is no longer the case for the dynamics of $h_0(\vec{r},\tau)$. Since any local operator conserves the the total height, the dynamics should preserve $\sum_{\vec{r}} h_0(\vec{r},\tau)$ at any given $\tau$, including the noise term $\zeta(\vec{r},\tau)$. As a result, we demand $\zeta(\vec{r},\tau) = \nabla \cdot \vec{\xi}(\vec{r}, \tau)$, where $\xi_{x,y}(\vec{r}, \tau)$ are two independent Gaussian white-noise field. Therefore, taken into account the rotation symmetry of the lattice, the noise $\zeta(\vec{r},\tau)$ satisfy,
\begin{equation}
    \begin{split}
        \langle \zeta(\vec{r},\tau) \rangle &= 0\\
        \langle \zeta(\vec{r},\tau) \zeta(\vec{r}', \tau') \rangle &= -2D \nabla^2 \delta^2(\vec{r} - \vec{r}') \delta(\tau - \tau'),
    \end{split}
    \label{eq:noise}
\end{equation}
which is of the same form as in volume conserving surface growth dynamics\cite{Sun1989PhysRevA.40.6763}.

\para Therefore, the Lagragian that give rise to the trajectory can be written as,
\begin{equation}
\begin{split}
        \mathcal{L} = & -(\partial_\tau h_0 + 2m h_0) \frac{1}{ D \nabla^2} (\partial_\tau h_0+ 2m h_0) \\
    &- \alpha \cos(2\pi h_0) - \beta (\nabla h_0)^2+ ...,
\end{split}
\end{equation}
where the $\alpha$, $\beta$ and $...$ terms denote the terms perturbing away from the RK point. The first term in the above Lagrangian reproduces the noise correlation given by Eq. \ref{eq:noise}.

\para Let us perform a Hubbard-Stratonivich (HS) decomposition of the first term and,
\begin{equation}
\begin{split}
       \mathcal{L}[\theta, h_0] = &i \theta (\partial_\tau h_0+2m h_0) + \frac{D}{4} (\nabla \theta)^2 \\&-\alpha \cos(2\pi h_0) - \beta (\nabla h_0)^2+ ...
\end{split}
\end{equation}
By dimensional analysis, the scaling dimension of $h_0$ is $3/2$, which makes a $h_0^2$ term marginal and a $(\nabla h_0)^2$ irrelevant. Near the minima of the cosine term, we can expand to the quadratic order in $h_0$, since higher orders are irrelevant. Then integrating out $h_0$ we have,
\begin{equation}
    \mathcal{L}[\theta] = \frac{1}{8\pi^2 \alpha }\left[(\partial_\tau - 2m)\theta\right]^2  + \frac{D}{4} (\nabla \theta)^2, 
\end{equation}
We then redefine $\theta_0 = \theta e^{-2 m \tau}$ and rescale the $\tau$ direction to get ,
\begin{equation}
    \mathcal{L}[\theta_0] = \frac{1}{8\pi^2 \alpha }\left(\partial_\tau\theta_0\right)^2  + \frac{D}{4}  (\nabla \theta_0)^2, 
\end{equation}
matching the effective action in the main text for the SN phase for small quantum fluctuation parametrized by $\alpha$.

From symmetry analysis, the potential term in $H_{\text{t-RK}}$ can be identified with the $\mathcal{L}_3$ term in Eq.\ref{eq:eff_lag} to the leading order. To get the SN phase, we need a flat configuration for the $\vec{\eta}$-fields, which requires $t > V$. The pinning of the $\vec{\eta}$ would then lead to the $\cos(2\pi h_0)$ term.

\end{document}